\documentclass[a4paper,11pt]{article}
\usepackage{amsmath,amssymb,color,graphics,epsfig}
\usepackage[text={16.3cm,24cm},centering]{geometry}

\usepackage{amsfonts}
\newcommand{\be}{\begin{equation}}
\newcommand{\ee}{\end{equation}}
\newcommand{\bea}{\setlength\arraycolsep{2pt} \begin{eqnarray}}
\newcommand{\eea}{\end{eqnarray}}
\newcommand{\nn}{\nonumber}
\def \mC {\mathcal C}
\def \mT {\mathcal T}
\def \cT {\mathcal T}
\def \la {\langle}
\def \ra {\rangle}
\def \hs {\hspace}

\def\ft#1#2{{\textstyle{\frac{\scriptstyle #1}{\scriptstyle #2} } }}
\def\fft#1#2{{\frac{#1}{#2}}}

\def\0{{\sst{(0)}}}
\def\1{{\sst{(1)}}}
\def\2{{\sst{(2)}}}
\def\3{{\sst{(3)}}}
\def\4{{\sst{(4)}}}
\def\5{{\sst{(5)}}}
\def\6{{\sst{(6)}}}
\def\7{{\sst{(7)}}}
\def\8{{\sst{(8)}}}
\def\sst#1{{\scriptscriptstyle #1}}

\def\D {\Delta}

\begin{document}
\title{\textbf{ Holographic Mutual Information of Two Disjoint Spheres}}
\author{Bin Chen$^{1,3,4}$,
%Peng-xiang Hao$^{1}$\footnote{pxhao@pku.edu.cn}~,
Zhong-Ying Fan$^{2}$,
 Wen-Ming Li$^{3}$ and
 Cheng-Yong Zhang$^{1}$\footnote{Email: bchen01,~liwmpku,~zhangcy0710@pku.edu.cn, fanzhy@gzhu.edu.cn}}
\date{}

\maketitle

\begin{center}

{\it $^{1}$Center for High Energy Physics, Peking University, No.5 Yiheyuan Rd, \\}
{\it Beijing 100871, P. R. China\\}
\smallskip
{\it $^{2}$ { Center for Astrophysics, School of Physics and Electronic Engineering, \\
 Guangzhou University, Guangzhou 510006, China }\\}
{\it $^{3}$Department of Physics and State Key Laboratory of Nuclear Physics and Technology,\\}
{\it Peking University, No.5 Yiheyuan Rd, Beijing 100871, P.R. China\\}
\smallskip
{\it $^{4}$Collaborative Innovation Center of Quantum Matter, No.5 Yiheyuan Rd,\\}
{\it  Beijing 100871, P. R. China\\}
\end{center}

\vspace{8mm}

\begin{abstract}
%\end{center}
We study quantum corrections to holographic mutual information for two disjoint spheres at a large separation by using the operator product expansion of the twist field. In the large separation limit, the holographic mutual information is vanishing at the semiclassical order, but receive quantum corrections from the fluctuations. We show that the leading contributions from the quantum fluctuations take universal forms as suggested from the boundary CFT. We find the universal behavior for the scalar, the vector, the tensor and the fermionic fields by treating these fields as free fields propagating in the fixed background and by using the $1/n$ prescription. In particular, for the fields with gauge symmetries, including the massless vector boson and massless graviton, we find that the  gauge parts in the propagators play indispensable role in reading the leading order corrections to the bulk mutual information.%, in spite of that the final results should be gauge independent.%. At leading order, we can compute the corrections for various quantum fluctuations using $1/n$ prescription. The bulk results are perfectly matched with the CFT results in the boundary.  We argue that this is due to the existence of the entangling surfaces, on which the gauge symmetry is effectively breaking. Hence, the mutual information of gauge fields at leading order can be interpreted as quantum entanglement of the new emergent degrees of freedom living on the two separated half-spheres associated with the breaking of gauge symmetry.
\end{abstract}

\baselineskip 18pt
\thispagestyle{empty}
\newpage

\thispagestyle{empty}

\pagebreak

\tableofcontents
\addtocontents{toc}{\protect\setcounter{tocdepth}{2}}

%%%%%%%%%%%%%%%%%%%%%%%%%%%%%%%%%%%%%%%%

%\newpage
%%%%%%%%%%%%%%%%%%%%%%%%%%%%%%%%%%%%%%%%

%\vspace{2cm}

\section{Introduction}

Entanglement is one of the most significant features of quantum physics, and plays an important role in understanding quantum many-body physics, quantum field theory, quantum information as well as quantum gravity. In quantum field theory, the entanglement entropy (EE) measures the entanglement  between an arbitrary subregion $A$ and its complement $\bar{A}$. It is defined as the von Neumann entropy of the reduced density matrix
\begin{equation}\label{eq:ee}
S_A=-\mathrm{tr}\rho_A  \log \rho_A
\end{equation}
where $\rho_A=Tr_{\bar{A}}\rho$ is the reduced density matrix of $A$ with respect to the density matrix of the whole system.  In practice, it is more convenient to compute the R\'enyi entropy first, which is defined as
\begin{equation}\label{eq:Renyi}
S_A^{(n)}=\frac{1}{1-n} \log\, \mathrm{tr} \rho_A^n,
\end{equation}
and then read the  entanglement entropy  by taking the limit
\begin{equation}
\lim_{n\to 1}S_A^{(n)}=S_A,
\end{equation}
provided that the continuation on $n$ is well-defined.
In quantum field theory, the computation of the R\'enyi entropy leads to the replica trick\cite{replicaHLW1994,replicaCalabrese2004,replicaCalabrese2005}
\be
\mathrm{tr}\rho_A^n=\frac{Z_n(C^n_A)}{Z^n}\,
\ee
where $Z_n$ and $Z$ are the partition functions of the theory on the conical spacetime $C_A^n$ and the original spacetime, respectively. The manifold $C_A^n$ comes from the identifications of the fields along the entangling surface. Equivalently one may introduce
the twist operators to induce the field identifications between different replicas, and consequently the partition function could be computed by the correlation functions of the twist operators in a replicated theory.  In general, it is difficult to compute entanglement entropy directly owing to the infinite degrees of freedom in a field theory.

In the past decade, the holographic entanglement entropy (HEE) has been studied intensively since its proposal in 2006 by Ryu and Takayanagi\cite{RT2006}. For a CFT dual to the Einstein AdS gravity, the entanglement entropy of the boundary subregion $A$ is given by the area of an  extremal surface $\gamma_A$ in the dual bulk
\begin{equation}\label{AL}
S_A=\frac{Area(\gamma_A)}{4G_N}.
\end{equation}
Here $G_N$ is the Newton constant and $\gamma_A$ shares the common boundary $\partial A$ with $A$ and is homologous to $A$. This so-called Ryu-Takayanagi (RT) formula is reminiscent of the Hawking-Bekenstein formula for the black hole entropy\cite{Bekenstein1973, BCH1973}. Actually, from the Euclidean gravity point of view, it has been proved that the holographic EE could be taken as a kind of gravitational entropy\cite{ML2013}, a generalization of the black hole entropy. The holographic entanglement entropy not only provides a new way to compute the entanglement entropy, but more importantly sheds new light on the holography and the AdS/CFT correspondence\cite{Maldacena:1997re,Gubser:1998bc,Witten:1998qj}. The various aspects on the holographic entanglement entropy can be found in the nice reviews\cite{Nishioka:2009un,Rangamani:2016dms}. %(\ref{AL}) is referred to the classical contribution and is often called the RT formula.

For two disjoint subregions $A$ and $B$, one can define the mutual information (MI) between them as
\begin{equation}
I(A, B)=S(A)+S(B)-S(A\cup B)\,.
\end{equation}
Different from the entanglement entropy, which is divergent in a field theory, the mutual information is free of ultraviolet(UV) divergence and is positive. It measures the entanglement between
two subregions: two entangled subsystems are correlated because they share an amount of information that is not foreseen classically. Actually, the mutual information satisfies an inequality\cite{Wolf}
\be
I({A,B})\geq \frac{\mC(M_A,M_B)^2}{2\parallel M_A\parallel^2 \parallel M_B\parallel^2},\label{inequality}
\ee
where $M_A$ and $M_B$ are the observables in the regions $A$ and $B$ respectively, and $\mC(M_A,M_B):=\la M_A\otimes M_B \ra-\la M_A \ra \la M_B\ra $ is the connected correlation function of $M_A$ and $M_B$. This indicates that the mutual information of two disjoint regions is usually not vanishing, even when the two regions are far apart, due to the quantum correlations between them.

Holographically, according to the RT formula, it is easy to verify that the holographic mutual information (HMI) suffers a phase transition from nonzero to zero when the separation distance $r>r_c$ \cite{Headrick:2010zt}, where $r_c$ is the critical distance. Thus, when considering the case that the two regions are far away enough, the holographic mutual information is simply vanishing. However, according to the inequality (\ref{inequality}), the holographic mutual information should not be vanishing. The discrepancy comes from the fact that the RT formula is given by the on-shell action of the gravitational configuration and only captures the leading order contribution to the entanglement entropy. After considering the quantum correction, the holographic mutual information is nonzero\cite{FLM2013,Barrella:2013wja}. In other words, the mutual information provides a nice probe to study the AdS/CFT correspondence beyond classical order. In particular, for the two-dimensional (2D) holographic CFT with a large central charge and sparse light spectrum, which is dual to the semiclassical AdS$_3$ gravity, the study of the  R\'enyi mutual information
allows us to read the 1-loop and even the 2-loop quantum corrections in gravity\cite{Chen:2013kpa,OPE}.

The direct computation of the mutual information is difficult since the replica trick leads to the conical geometry which could be not only of singularity but also of nontrivial topology. For example, in two dimensions, the pasting of the multi-intervals leads to a higher genus Riemann surface. Nevertheless, when two disjoint regions are far apart, one may use the operator product expansion(OPE) of the twist operators to compute the large distance expansion of the (R\'enyi) mutual information. This turns out to be quite effective for 2D CFT\cite{Headrick:2010zt,Chen:2013kpa,Calabrese:2010he}. It can actually be applied to the higher dimensional case as well. In \cite{Cardy:2013nua}, the leading order mutual information of the disjoint spheres for free scalars has been discussed by using the OPE of spherical twist operator\cite{Hung:2014npa} and found to be consistent with the numerical results\cite{Shiba:2012np,Casini:2008wt}. The discussion has been generalized to the next-to-leading order mutual information in \cite{Agon:2015twa} and the R\'enyi mutual information in \cite{Chen:2016mya,Long:2016vkg} for free scalars.

It is definitely interesting to have a better understanding of the mutual information in a general CFT, beyond the free scalar theories. At the first sight this turns out to be a formidable problem, because even for the simplest two-sphere case  the computation in the OPE of the twist operator involves the one-point functions of the primary operators in the conical geometry, which requires the detailed information of the CFT. Therefore it is really surprising to find that the mutual information of two disjoint spheres presents universal behaviors at the first few leading orders\cite{CCHL2017}. For a generic CFT, it was further proposed in \cite{CCHL2017} that the mutual information could be expanded in terms of the conformal block
\be
I({A,B})=\sum_{\D,J}{ b_{\D,J}}{ G_{\D,J}(u,v)},
\ee
where $\D$ and $J$ are the conformal dimension and the spin of the primary operator propagating between two spheres, and $G_{\D,J}$ is the conformal block. As the conformal block in the diagonal limit could be approximated by
\be
G_{\D,J}(z) \simeq z^\D+ \cdots,
\ee
the contributions are dominated by the operators of the first few lowest dimensions. The remarkable point is that the coefficients $b_{\D,J}$ take universal forms for the operators giving dominant contributions. For example, for a CFT in which the primary operator $\mathcal{O}$ of the lowest dimension $\D$ is a scalar type, then the leading contribution comes from the bilinear operators $O^{(s)(j_1j_2)}=\mathcal{O}^{(j_1)}\mathcal{O}^{(j_2)}$ ($j_i$  labels the replica and superscript $(s)$ stands for operators constructed from scalar type operators) with the coefficient\cite{AF2015}
\be
b^{(s)}_{2\Delta,0}=\frac{\sqrt{\pi}\Gamma[2\Delta+1]}{4^{2\Delta+1}\Gamma[2\Delta+\frac{3}{2}]},
\ee
while the next-to-leading one could be from the bilinear operators with a derivative\footnote{Strictly speaking, whether or not this is the operator giving the next-to-leading contribution depends on the spectrum of the theory and the dimension $\D$. Here we assume that the operator of the next lowest dimension is of dimension at least $1/2$ higher and the lowest operator is of the dimension greater than $1/2$ as well.}
\be
O^{(s)(j_1j_2)}_{\mu}=\mathcal{O}^{(j_1)}\partial_{\mu}\mathcal{O}^{(j_2)}-(j_1\leftrightarrow j_2),
\ee
with the coefficient
\be
b^{(s)}_{2\Delta+1,1}=-\frac{\sqrt{\pi}\Delta\Gamma[2\Delta+1]}{2^{4\Delta+3}\Gamma[2\Delta+\frac{5}{2}]}.
\ee
The coefficients $b$'s are independent of the OPE coefficients of the theory. These universal behaviors persist no matter the operator of the lowest dimension is fermionic, vector or tensor type. Actually, one needs to know the exact spectrum of the CFT in order to know the leading contributions to the mutual information. Once the spectrum of the CFT is known, for example by using the bootstrap techniques, the leading contributions can be read. 

In this paper, we would like to understand these universal behaviors in a holographic way.
In \cite{FLM2013}, Faulkner, Lewkowycz  and Maldacena (FLM) proposed that the quantum corrections to the HEE are essentially given by the bulk entanglement entropy between the bulk region $A_b$ enclosed by $\gamma_A \cup A$ and its complement $\bar{A_b}$. %That is to say, to the next leading order, one can take the bulk field  living on a fixed background geometry. %The EE can be calculated just as what we usually do in ordinary quantum field theory.
While this proposal gives us a prescription for calculating the quantum corrections to the entanglement entropy, it is technically challenging to carry out such computations. One technical difficulty is that the bulk geometry corresponding to the replicated geometry is hard to determine due to the large backreaction\cite{ML2013,Dong2016}.  However, for the holographic mutual information we are interested in, the backreaction can be ignored. Consequently one can compute the mutual information holographically by using the OPE of the twist operators.
 Just like other non-local  Wilson-line and Wilson loop operators\cite{Berenstein:1998ij,Gomis:2009xg,Chen:2007zzr}, the OPE of the twist operator can be computed in a holographic way. In \cite{AF2015}, Ag\'on and Faulkner computed the leading order mutual information coming from scalar field holographically and found agreement with the field theory result.  In this work, we study the quantum corrections to the holographic MI in more general cases, including the higher order contributions coming from the scalar field and the leading order  contributions coming from non-scalar fields, including the massless vector boson, the massless graviton, the fermion and also the massive fields. We reproduce the universal behaviors found in \cite{CCHL2017} exactly.

%We carry out computations both in boundary CFT and in the bulk. We find that the bulk results are perfectly matched with the CFT results on the boundary. In particular, for gauge fields including vector bosons and gravitons, the pure gauge parts, instead of the physical parts, in the propagators gives the non-trivial leading order corrections to mutual information, in spite of that the final results should be gauge independent.

 The remaining parts of this paper are organized as follows. In the next section, after giving a brief review of the spherical twist operator and its OPE expansion, we introduce the field theory computation on the  mutual information from scalar, vector and tensor type operators in CFT\cite{CCHL2017}. In section 3, we investigate the bulk computation. By doing the operator product expansion of the extremal surface operator, we get the quantum corrections of the scalar, the gauge boson, the graviton and the fermion to the holographic mutual information. Especially we find that the gauge part of propagators of the massless vector boson and massless graviton play an important role in the computation.  We end with conclusions and discussions in section 4. In the appendices, we collect our computations on the massive vector boson and massive graviton, and also the formulae on the graviton propagator.  Without confusion, we work in the Euclidean signature throughout this paper.

 \section{Field theory results}

 Let us consider the mutual information of two disjoint spheres in a $d$-dimensional CFT. By using the global conformal symmetry, we can always set the radii of two spheres to be $R$ and the centers of two spheres to be one at the origin and the other at $x_1=1, x_i=0, i\geq 2$ respectively.
 Now the only independent conformal invariant quantity is the cross ratio
\be
z=\bar{z}=4R^2,\hs{5ex}u=z\bar{z},\hs{3ex}v=(1-z)(1-\bar{z}).\nn
\ee
In the disjoint case, we have $0<z<1$.

 We would like to compute the mutual information of two disjoint spheres. The mutual information is given by
\begin{equation}
I(A, B)=\lim_{n\to 1}I^{(n)}(A, B)=-\lim_{n\to 1}\frac{1}{1-n}\log\left(\frac{Z_n(C_{A\cup B}^{(n)})Z^{n}}{Z_n(C_{A}^{(n)})Z_n(C_{B}^{(n)})}\right)\,,\label{eq:RenyiMI}
\end{equation}
where $I^{(n)}(A, B)$ is the R\'enyi mutual information.  The partition functions  can be calculated using the nonlocal twist operators $\mT^{(n)}$.
\begin{equation}
\frac{Z_n(C_{A\cup B}^{(n)})}{Z^{n}} =
\left\langle \mT_{A}^{(n)}\mT_{B}^{(n)}\right\rangle _{M^{n}}.
\end{equation}
 Here  $M^n$ stands for $n$ copies of the original space. $\mT_{A}^{(n)}$  and $\mT_{B}^{(n)}$ stand for nonlocal twist operators corresponding to the regions $A$ and $B$ respectively.
 In the large distance regime, we can treat the twist operator as a semi-local operator. It can be expanded in terms of the primary operators of the replicated theory
\be
\cT^{(n)}=<\cT^{(n)}>\sum_{\{\D,J\}}c_{\D,J}Q[O_{\D,J}],
\ee
where $Q[O_{\D,J}]$ denotes all the operators generated from the primary operator $O_{\D,J}$ of dimension $\D$ and spin $J$. Note that the summation is over all the primary operators in the $n$-replicated CFT. The coefficient $c_{\D,J}$ is read from
 the one-point function of the primary operator in the presence of the spherical twist operator. Equivalently it can be computed by the one-point function of the primary operator in the conical geometry. In 2D, the coefficient can be read by using the uniformization map. In higher dimensions, it is difficult to compute, except the case that the theory is free such that one can use  the method of images.

 The R\'enyi mutual information is captured by
 \bea
 \frac{ \la \cT^{(n)}_A \cT^{(n)}_B\ra}{ \la \cT^{(n)}_A \ra   \la \cT^{(n)}_B \ra}&=& \sum_{\{\D,J\}}c^2_{\D,J}\la Q_A [O_{\D,J}]Q_B [O_{\D,J}] \ra\nn\\
 &=&\sum_{\{\D,J\}}s_{\D,J}G_{\D,J}(u,v).
 \eea
 where the building block is the two-point function of the primary module, the conformal block\cite{Dolan:0011,Dolan:0309}. The coefficient $s_{\D,J}$ is given by
\be
s_{\Delta,J}=f_{\Delta,J}\sum_{O_{\Delta,J}}\frac{a^2_{\Delta,J}}{N_{\Delta,J}}, \label{sDJ}
\ee
where the summation is over all the primary operators with the same $(\D,J)$ in the replicated theory, $a_{\D,J}$ is determined by the one-point function of the operator $O_{\D,J}$ in the planar conical geometry
\be
\la O_{\D,J}(x)\ra_n=a_{\D,J}\frac{T_J}{|x|^\D}
\ee
with $T_J$ being a kind of tensor structure. $N_{\D,J}$ in (\ref{sDJ}) is the normalization factor in the two-point function in the flat spacetime.
\be
\langle{O}_{\Delta,J}(x){O}_{\Delta,J}(x^{\prime})\rangle=N_{\Delta,J}\frac{T^{\prime}_{J}(x-x^{\prime})}{(x-x^{\prime})^{2\Delta}}
\ee
with $T'_J$ being the tensor structure relating to the operator with spin $J$.
The coefficient $f_{\D,J}$ could be determined by considering one spherical operator and mapping it to a half plane. It depends only on the tensor structure of the operator\footnote{For the detailed study on the tensor structure and the coefficient $f$, please refer to \cite{CCHL2017}.}. In terms of the conformal blocks, the R\'enyi mutual information can be expressed by \be
I^{(n)}({A,B})=-\frac{1}{1-n}\log (1+\sum_{\{\D,J\}}s_{\D,J}G_{\D,J}),
\ee
and the mutual information is just
\be
I({A,B})=\sum_{\{\D,J\}}b_{\D,J}G_{\D,J}(u,v) ,
\ee
with the coefficient $b_{\D,J}$ being related to the expansion of $s_{\D,J}$ in powers of $(n-1)$
\be
b_{\D,J}=\frac{\partial s_{\D,J}}{\partial n}\big|_{n=1}.
\ee
This is the conformal block expansion of the mutual information. As the conformal block in the diagonal limit is approximated by\cite{Matthijs:1305}
\be
G_{\D,J}(z) \simeq z^\D+ \cdots,
\ee
the leading contribution to the mutual information is from the primary operator with the lowest dimension and nonvanishing coefficient.

As the one-point functions of the operators purely in one replica is simply zero, they  give vanishing mutual information. It turns out that the dominant one comes from the bilinear operators composed of the operators in different replicas. For example, for a CFT in which the primary operator $\mathcal{O}$ of the lowest dimension $\D$ is of scalar type, then the leading contribution comes from the bilinear operators $O^{(s)(j_1j_2)}=\mathcal{O}^{(j_1)}\mathcal{O}^{(j_2)}$ ($j_1\neq j_2$). The next-to-leading one comes from the bilinear operators with a derivative
\be
O^{(s)(j_1j_2)}_{\mu}=\mathcal{O}^{(j_1)}\partial_{\mu}\mathcal{O}^{(j_2)}-(j_1\leftrightarrow j_2).\label{Jscalar}
\ee
One important point is that the primary operators of the replicated theory could be not just the tensor products of the operators $\mathcal{O}^{(j)}$ in different replicas. The above bilinear operator with a derivative  is a typical example. This shows that the spectrum of the replicated theory is much involved. Even for the free scalar, there is no systematical way to construct the primary operators in the replicated theory\cite{Chen:2016mya}. However, if we are satisfied with the leading contributions to the mutual information, the relevant operators can be constructed explicitly\footnote{Note that for the 2D orbifold CFT, there could be other ways to construct the spectrum in terms of the operators in the twist sectors rather than the normal sector, depending on how to insert the complete set of state basis\cite{Chen:2014hta,Chen:2014ehg,Chen:2015kua}.}.

 %Composite operators $O^{(j)},O^{(jj')}$ and the corresponding OPE coefficients $C_{(j)}^A,C_{(jj')}^A$ can have spin indexes. We just write it in a short-cut way. Indexes $j,j'$ indicate operator insertions of $O$ on different replicas $j,j'=0,...,n-1$. The perfactors on the right hand side is inserted since we expect the leading behavior at large distance regime to come from the term when all the operators $O$ are the identity operators, and in this limit $\mT_{A \cup B}^{(n)} \sim \mT_A^{(n)}\mT_B^{(n)}$. Assuming a $Z_2$ symmetry for the lowest dimension operators such that the symmetry is not spontaneously broken in the replica manifold,  the operators located on just one sheet have vanishing contributions to the MI \cite{AF2015}. Thus, the leading contribution comes from the two-sheet operators with lowest scaling dimensions. As a result, when considering large separation limit, the MI can be calculated as

It is remarkable that the coefficients for the leading contributions take  universal forms, which means that they depend only the scaling dimensions and the spins of the primary operators and have nothing to do with the construction of the CFT itself. Naively one cannot expect to get such  universal behaviors as the one-point function of the primary operator in a conical space cannot be determined in a simple way. It is feasible because the one-point function get simplified in the $n\to 1$ limit. This leads to the so-called $1/n$ prescription\cite{CCHL2017}.

The cause of the $1/n$ prescription is as follows. Let $G_n$ be any periodic function in the conical geometry whose angular direction is identified as $\theta \sim \theta + 2\pi n$. It satisfies
\begin{equation}
G_n(r,\theta,y^i) = G_n(r,\theta + 2\pi n,y^i).
\end{equation}
In the limit $n\rightarrow 1$, it returns to the usual function on the original flat space $G_1(r,\theta,y^i) = \lim_{n\rightarrow 1} G_n(r,\theta,y^i)$.  Using the Fourier expansion, one can show that
\begin{equation}
G_n(r,\theta,y^i) = G_1(r,\theta /n,y^i)+{O}(n-1).
\end{equation}
 The periodic function in the conifold in the limit $n\rightarrow 1$ is related to the function in the original space by dividing the angular variable by $n$. This is called the $1/n$ prescription\cite{CCHL2017}. Consequently, we have
\begin{equation}
\lim_{n\rightarrow 1}\sum_{q=1}^{n-1} \frac{G_n^2(2\pi q)}{n-1} = \lim_{n\rightarrow 1}\sum_{q=1}^{n-1} \frac{G_1^2(2\pi q/n)}{n-1}.
\end{equation}
This is very useful for calculating the expansion coefficients $b_{\D,J}$.

For the examples we discussed before, the coefficient of the bilinear operator $O^{(s)(j_1j_2)}$ turns out to be
\be
b^{(s)}_{2\Delta,0}=\frac{\sqrt{\pi}\Gamma[2\Delta+1]}{4^{2\Delta+1}\Gamma[2\Delta+\frac{3}{2}]},\label{coeffscalar}
\ee
while the coefficient for the operator (\ref{Jscalar}) is
\be
b^{(s)}_{2\Delta+1,1}=-\frac{\sqrt{\pi}\Delta\Gamma[2\Delta+1]}{2^{4\Delta+3}\Gamma[2\Delta+\frac{5}{2}]}.\label{coeffJs}
\ee
The coefficient (\ref{coeffscalar}) was first derived in
 \cite{AF2015}. %Strictly speaking, the method of images is only applicable for the free theory to compute the one-point function of the operator in a conical space. In an interacting theory, one has to take into account of the interaction in the computation. However, to the leading order of $(n-1)$, the method of image can lead to the same answer as the $1/n$ prescription. %In the next section, we will introduce the field theory result and show that the method of images

% the authors computed mutual information between two spheres at large separation limit for both conformal field theories living in flat spacetime and corresponding bulk theories living in AdS spacetime independently. The results are coincident. They used the image method to calculate the one-point function in conifold. Strictly, image method is only applicable for free theory.

%To verify FLM proposal \cite{FLM2013} further and to remedy this flaw, we try to compute the MI in more general cases using the $1/n$ prescription in this work. We will calculate the leading contributions coming from the bulk gauge vector fields and gravitons, and also the massive vector and tensor fields. Due to the gauge parts in the propagators and other uncertainties, this is a non-trivial task. We will get a clearer cognition of various operators/ fields correspondence.

%%%%%%%%%%%%%%%%%%%%%%%%%%%%%%%%%%%%%%%%%%%%%%%%%%%%%%%%%%%%%%%%%%%%%%%%%%%%%%%%%%%%%%%%%%%%%%%%%%%%%%%%%%%%%%%%%%%%%%%%%%%%%%%%%%%%%%%%%%%%%%%%%%%%%%%%%%%%%%%%%%%%%%%%%%%%%%%

%\section{CFT$_d$ in Euclidean spacetime}
\subsection{Scalar type operator}

Let us first review the calculations of the leading order contribution to the mutual information from a primary scalar operator in the boundary CFT by using  the $1/n$ prescription. In this case, the operator giving the leading order contribution is of the type $O^{(s)(jj')}=\mathcal{O}^{(j)} \mathcal{O}^{(j')}$ where  $\mathcal{O}^{(j)}$ is a scalar primary operator with the lowest dimension $\Delta$ living on the $j$-th replica. For this operator, its two-point function at the leading order in the large distance limit is given by
 \begin{equation}
 \langle O^{(s)(0j)} (x_A)^{(s)(0j)} (x_B)\rangle_{M^n}\simeq \frac{1}{|x_A-x_B|^{4\Delta}}\,,
 \end{equation}
where we have used the two-point function on the plane in CFT,
\begin{equation}\label{eq:2pointscalar}
\langle \mathcal{O}^{(j)}(x_A)\mathcal{O}^{(j)}(x_B)\rangle_M=\frac{1}{|x_A-x_B|^{2\Delta}}.
\end{equation}
%We shall point out that from (\ref{eq:MI}) and (\ref{eq:coef}), the normalization constant of the two-point function is irrelevant in the mutual information at leading order. This is also true for higher spin operators although the situations in those cases are more complicated.

In order to compute the OPE coefficients, we do a conformal transformation
\be
x'^\mu=\fft{x^\mu+c^\mu}{\Omega}-\fft{c^\mu}{2|c|^2}\,,\quad \Omega=c^2+2\,c\cdot x+x^2\,,\label{ct}
\ee
where $c^\mu$ is a $d$-dimensional constant vector, given by $c^\mu=(0,R,0,\dots,0)$. Under this transformation, the original conifold geometry ${C}_A^{(n)}$ with spherical conical singularity ($t_{\mathrm{E}}=0\,,\delta_{ij}x^i x^j=R^2$) is conformally transformed into a new
conifold geometry ${C}_A^{'(n)}$ with the singularity located at the plane ($t'_{\mathrm{E}}=0\,,x'^1=0$). Moreover, the infinity  $x_{\infty}=(t_{\mathrm{E}}=0\,,x^i_\infty)$ is transformed into a finite point
$x'_{\infty}=(t'_{\mathrm{E}}=0\,,x'^i=-\fft{c^i}{2c^2})$. The Jacobian is approximately (note that $\Omega\mid_{x\rightarrow \infty}\approx |x|^2$)
\be \fft{\partial x'^\mu}{\partial x^\nu}\approx\Omega^{-1} I^\mu_\nu(x)\,,\qquad I_{\mu\nu}(x)=\delta_{\mu\nu}-2\hat{x}^\mu\hat{x}^\nu  \,,\ee
where $\hat{x}^\mu=x^\mu/|x|$ is a unit vector. Under the transformation, we find that at the infinity  a general spin-$s$ primary operator with the scaling dimension $\Delta$ transforms as
\begin{equation}
\label{spin} \langle O_{\mu_1\mu_2\dots \mu_s}(x_\infty)\rangle_{{C}_A^{(n)}}\approx \Omega^{-\Delta}(x_\infty)I^{\nu_1}_{\mu_1}(x_\infty)I^{\nu_2}_{\mu_2}(x_\infty)\dots I^{\nu_s}_{\mu_s}(x_\infty)\langle O_{\nu_1\nu_2\dots \nu_s}(x'_\infty)\rangle_{{C}_A^{'(n)}} \,.
\end{equation}
It should be emphasized that the right-hand side of this equation is only the leading term, which however is sufficient for us to calculate the OPE coefficients. Then for a scalar operator, we find
\begin{equation}
 C^A_{(jj')}=\langle \mathcal{O}^{(j)}(x'_\infty)\mathcal{O}^{(j')}(x'_\infty) \rangle_{{C}_A^{'(n)}}\,.
 \end{equation}
The one-point function on the new conifold geometry ${C}_{A}^{'(n)}$ can be computed using two different methods, as had been done in \cite{AF2015} and \cite{CCHL2017}. For a general CFT, the coefficient is theory-dependent. In \cite{AF2015}, it was shown that the correlators in the conical space  could be transformed to the correlators on the hyperbolic space at finite temperature via the map suggested by H. Casini et.al. in \cite{Casini:2011kv}. Moreover by using the 
analyticity and the properties of the thermal field theory, the authors of \cite{AF2015} read the contribution from the bilinear operator to the mutual information 
\begin{equation}
\label{smi} I(A\,,B)=\fft{\sqrt{\pi}\,\Gamma(2\Delta+1)}{4^{2\Delta+1}\Gamma(2\Delta+3/2)}z^{2\Delta} \,.
\end{equation}
This is the leading contribution of a scalar operator with the scaling dimension $\Delta$ to the mutual information.

Now we would like to use the $1/n$ prescription to derive the same result. As proposed in \cite{CCHL2017}, the Green function $G_n(\theta)$ with a period $2\pi n $ living on the conifold can be expanded as $G_n(\theta)=G_1({\theta}/{n})+{O}(n-1)$
where we suppose $G_n$ is analytically continuable with $n$, and $\lim_{n\to 1}G_n(\theta)=G_1(\theta)$. In other words, when $n$ is close to unity, the Green function $G_n(\theta)$ at the leading order on the conifold geometry $C_A^{(n)}$ is simply given by its counterpart on the plane with the angle coordinate $\theta $ divided by $n$. For the bilinear scalar operators, we have
\begin{equation}
C_{(jj')}^A\mid_{n\to 1}=\lim_{n\to 1}\langle \mathcal{O}^{(j)}(x'_\infty)\mathcal{O}^{(j')}(x'_\infty)\rangle_{{C'}_A^{(n)}}=\lim_{n\to 1} R_A^{2\Delta}\mathrm{sin^{-2\Delta}\big(\ft{\theta_j-\theta_{j'}}{2n}\big)}\,.
\end{equation}
Substituting the above formula into the mutual information, we get
\begin{equation}
I(A,B)=\lim_{n\to1}\frac{n}{2^{4\Delta+1}(n-1)}\sum_{j=1}^{n-1}\sin^{-4\Delta}\big(\frac{j\pi}{n}\big) z^{2\Delta}\,,
\end{equation}
where we have set $\theta_j=2\pi j$. Provided the equality
\begin{equation}
\sum_{j=1}^{n-1}\sin^{-\Delta}\frac{j\pi}{n}=(n-1)\frac{\sqrt{\pi}}{2}\frac{\Gamma(\frac{\Delta}{2}+1)}{\Gamma(\frac{\Delta}{2}+3)}\,,
\end{equation}
we immediately arrive at the same answer (\ref{smi}).

%\subsection{Image method vs 1/n prescription: general discussions}

The essence of the $1/n$ prescription is that the Green's function in a conical geometry could be approximated by the Green's function in a flat spacetime in the expansion by the orders of $(n-1)$. In the leading order of $(n-1)$, the Green's function is directly related to the one in the flat spacetime. For the bilinear operator, its one-point function in the conical geometry is well approximated by the two-point function of single operators. This  suggests that in the leading order of $(n-1)$ the operator $\mathcal{O}$ could be approximated to be the one in free CFT without considering the interaction. Actually, if one naively take the operator as a generalized free field, one can get the above result by using the method of images which is only applicable in the free theory. In other words, to the leading order the relevant operators could be taken as the ones in a generalized free theory.

%reduces exactly to the method of images. This is why the  method of images can give the same answer as the $1/n$ prescription, even though it is only applicable in the free theory. 
%\subsection{Scalar operator: other contributions}

To compare with the bulk computation in the next section, we list the other contributions from the operators in the replicated theory composed of the scalar operator in the mother CFT\footnote{The detailed computation on the coefficients of the bilinear operators constructed from the scalar, the vector  and the tensor type operator from mother CFT,  can be found in \cite{CCHL2017}. }. Besides the bilinear one  and the spin-1 one discussed before, there are other types of operators. The next one is the spin-2 operator, defined by
\be
O^{(s)(j_1j_2)}_{\mu\nu}=\frac{1}{2}\mathcal{P}^{\alpha\beta}_{\mu\nu}(\partial_{\alpha}\mathcal{O}^{(j_1)}\partial_{\beta}\mathcal{O}^{(j_2)}
-\frac{\Delta}{\Delta+1}\mathcal{O}^{(j_1)}\partial_{\alpha}\partial_{\beta}\mathcal{O}^{(j_2)}+(j_1\leftrightarrow j_2)), \label{tscalar}
\ee
where 
\be
\mathcal{P}^{\alpha\beta}_{\mu\nu}=\frac{1}{2}\left(\delta_{\mu}^{\alpha}\delta_{\nu}^{\beta}+\delta_{\nu}^{\alpha}\delta_{\mu}^{\beta}\right)-\frac{1}{d}\delta_{\mu\nu}\delta^{\alpha\beta}
\ee
 is the operator projecting the tensor to its symmetric and traceless part. Its coefficient in the mutual information is\cite{CCHL2017}
\be b^{(s)}_{2\Delta+2,2}=\frac{(d-1)\sqrt{\pi}(2+4\Delta+3\Delta^2)\Gamma[2\Delta+3]}{d(2\Delta+1)^22^{4\Delta+5}\Gamma[2\Delta+\frac{7}{2}]}.\label{b2dp22}
\ee
%%%%%%%%%%%%%%%%%%%%%%%%%%%%%%%%%%%%%%%%%%%%%%%%%%%%%%%%%%%%%%%%%%%%%%%%%%%%%%%%%%%%%%%%%%%%%%%%%%%%%%%%%%%%%%%%%%%%%%%%%%%%%%%%%%%%%%%%%%%%%%%%%%%%%%%%%%%%%%%%%%%%%%%%%%%%%%

\subsection{Vector type operator}

If the operator of the lowest dimension $\D$ in the mother CFT is a vector type $J^\mu$, then the bilinear operator giving the leading contribution to the mutual information could be of the following forms
\be
O^{(v)(j_1j_2)}_{\mu\nu}=\mathcal{P}^{\alpha\beta}_{\mu\nu}(J_{\alpha}^{(j_1)}J_{\beta}^{(j_2)}),\hs{3ex} O^{(v)(j_1j_2)}=J_{\mu}^{(j_1)}J^{(j_2)\mu}.
\ee
The superscript $(v)$ stands for operators constructed from vector type operators. Their coefficients in the expansion of the mutual information are respectively\cite{CCHL2017}
\bea
b^{(v)}_{2\Delta,0}&=&\frac{(d-2)^2}{d}\frac{\sqrt{\pi}\Gamma[2\Delta+1]}{4^{2\Delta+1}\Gamma[2\Delta+\frac{3}{2}]},\label{vector0}\\ b^{(v)}_{2\Delta,2}&=&\frac{(d-1)}{d}\frac{\sqrt{\pi}\Gamma[2\Delta+1]}{4^{2\Delta}\Gamma[2\Delta+\frac{3}{2}]}.\label{vector2}
\eea

\subsection{Tensor type operator}

The construction can be generalized to other types of tensor operator. Here we only consider the symmetric spin-2 operator. One typical example of such type is the stress tensor, which satisfies the conservation law. We denote the spin-2 tensor as $T_{\mu\nu}$ but do not requires it to be a stress tensor. Suppose the spin-2 operator  is the operator of the lowest dimension $\D$ in the mother CFT. Its bilinear form can be decomposed  into six classes, among which only three of them have nonvanishing contribution to the mutual information. They are of the following forms respectively
\bea
&&O^{(t)(j_1j_2)}=T_{\mu\nu}^{(j_1)}T^{(j_2)\mu\nu},\nn\\ &&O^{(t)(j_1j_2)}_{\mu\nu}=\mathcal{P}^{\alpha\beta}_{\mu\nu}(T_{\alpha\gamma}^{(j_1)}T_{\beta}^{(j_2)\gamma}),\nn\\
&&O^{(t)(j_1j_2)}_{\mu\nu\rho\sigma}=\mathcal{P}^{\alpha\beta\gamma\delta}_{\mu\nu\rho\sigma}(T_{\alpha\gamma}^{(j_1)}T_{\beta\delta}^{(j_2)}),
\eea
where the superscript $(t)$ stands for operators constructed from tensor type operators. $\mathcal{P}^{\alpha\beta}_{\mu\nu}$ is a projection operator defined in (\ref{tscalar}) and
\be \mathcal{P}_{\mu\nu\rho\sigma}^{\alpha\beta\gamma\delta}= \frac{1}{24}\delta_{(\mu}^{\alpha}\delta_{\nu}^{\beta}\delta_{\rho}^{\gamma}\delta_{\sigma)}^{\delta}-\frac{1}{12(d+4)}\delta_{(\mu\nu}\delta^{(\alpha\beta}\delta_{\rho}^{\gamma}\delta_{\sigma)}^{\delta)}+\frac{1}{3(d+2)(d+4)}\delta_{(\mu\nu}\delta_{\rho\sigma)}\delta^{(\alpha\beta}\delta^{\gamma\delta)}.
\ee
 Their coefficients in the conformal block expansion of the mutual information are respectively\cite{CCHL2017}
\bea
b^{(t)}_{2\Delta,4}&=&\frac{\sqrt{\pi}(d^2-1)\Gamma[2\Delta+1]}{(8 + 6 d + d^2)2^{4\Delta-2}\Gamma[\frac{3}{2} + 2 \Delta]},\label{tensor4}\\
b^{(t)}_{2\Delta,2}&=&\frac{\sqrt{\pi}(d-2)(d-1)\Gamma[2\Delta+1]}{(d+4)2^{4\Delta}\Gamma[\frac{3}{2} + 2 \Delta]},\label{tensor2}\\
b^{(t)}_{2\Delta,0}&=&\frac{\sqrt{\pi}(d-2)^2(d-1)\Gamma[2\Delta+1]}{(d+2)2^{4\Delta+3}\Gamma[\frac{3}{2} + 2 \Delta]}.\label{tensor0}
\eea

\section{Bulk mutual information}

\begin{figure}[h]
\begin{centering}
\includegraphics[scale=0.7]{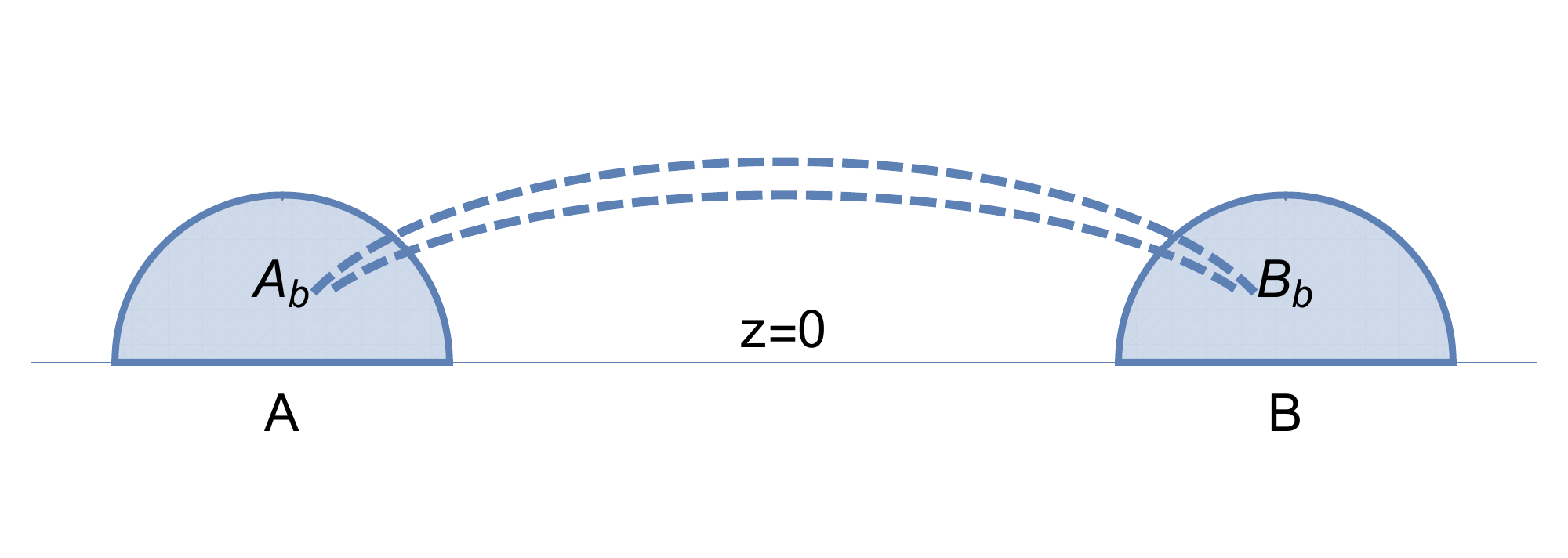}
\par\end{centering}
\caption{\label{fig:HMIDigram} The mutual information between two boundary regions $A$ and $B$ can be computed holographically by exchanging the bulk fields. Here the distance between $A$ and $B$ is much larger than their radius. $A_b$ and $B_b$ are bulk regions enclosed by the corresponding minimal surfaces.}
\end{figure}

In \cite{FLM2013}, it was argued that quantum corrections to the holographic entanglement entropy are essentially given by the bulk entanglement entropy between the subregion enclosed by the RT surface and its complement. We refer this the FLM proposal.  It is hard to test the FLM proposal since the bulk computations of the entanglement entropy are in general very difficult. Fortunately, according to the FLM proposal, in the long distance regime the MI of two disjoint boundary subregions equals to the bulk MI between the corresponding two bulk subregions surrounded by the RT surfaces and the boundary, as shown in Fig. \ref{fig:HMIDigram}. In particular,  the bulk MI for two hemispheres can be analytically computed by adopting the OPE technique. This was first done in \cite{AF2015} for a free scalar field at the leading order in the large distance limit. The results support the FLM proposal. In this section, we first extend the study of a free scalar field to the next-to-leading order and the next-to-next-to-leading order. This is nontrivial since we need carefully construct the gravity duals of the  primary operators at different replicas for the boundary CFT. We further calculate the bulk MI coming from the gauge boson, the graviton and the fermion. In all these cases, our bulk results are well matched with the CFT results reported in \cite{CCHL2017} and hence verify the FLM proposal in a great careful manner.

When adopting the OPE method in the bulk, we immediately encounter a difficult problem. The gravity dual of the R\'enyi entanglement entropy (a modular version) is one quarter of the area of a cosmic brane with the tension \cite{ML2013,Dong2016}
\begin{equation}
T_n=\frac{n-1}{4nG_N}\,,
\end{equation}
which is anchored on the boundary. If $n\neq 1$, the cosmic brane is heavy and would change the spacetime. Consequently one has to solve the equations of motion of the gravity coupled with the cosmic brane. Technically speaking, this is very difficult to handle, even in numerical ways\footnote{In the semi-classical AdS$_3$/CFT$_2$, one can extend the Schottky uniformization into the bulk to find the gravitational configuration dual to the higher genus Rieman surface. The gravitational configurations are not the minimal surfaces\cite{Krasnov:2000zq,Hartman:2013mia,Faulkner:2013yia}.}. Fortunately, our goal is to compute the bulk MI rather
than the general R\'enyi MI. This only requires us to consider a sufficiently light cosmic brane as $n$ close to unity. In this case, the cosmic brane becomes effectively tensionless such that we can work in the probe limit and ignore the backreaction. As a result, we can still treat a spherical twist operator as a hemisphere in the bulk ignoring the deformation. In other words, the holographic description of the spherical twist operator is a nonlocal hemisphere in the bulk. Moreover, each hemisphere can be described by the operator product expansion. This is similar to the holographic description of the Wilson loop or surface operator and its OPE\cite{Berenstein:1998ij,Chen:2007zzr}.

As we argued above, the holographic configuration corresponding to the sphere is a hemisphere. This is only true when we take the $n\to 1$ limit which suggest that the dual configuration is a RT surface. However, when we apply the replica trick, the boundary sphere becomes a conical space such that the dual configuration should be very different. Nevertheless, as we are going to take $n\to 1$ limit, we expect that the bulk configuration is well-approximated by the hemisphere with transverse direction being a conical space. Simply speaking, the bulk configuration is approximated by a replicated geometry as well. Such a holographic twist operator can be expanded as
\be
\cT^{(n)}_A \sim 1+\sum_{j=0}^{n-1}C_{(j)}^AO^{(j)}(x_A)+\sum_{j\neq j'}C_{(jj')}^AO^{(jj')}(x_{A})+...\,,
\ee
where  the normalization factor has been dropped and $C$'s are the expansion coefficients. The operator $O^{(j)}$ stands for the operator at one replica, while the operator $O^{(jj')}$ stands for the operator composed of the operators in two different replicas, etc.. Note that these operators may have nonzero spin. The expansion coefficients can be read from the one-point functions of the operators in the presence of the twist operator. Actually the one-point function of $O^{(j)}$ is vanishing, and the first non-trivial one is the bilinear operator constructed from the fields in two different replicas. For example, from the scalar field $\phi$ dual to the scalar operator, there is
 a bulk bilinear operator $O^{(s)(jj')}=\phi^{(j)}\phi^{(j')}$. Its one-point OPE takes the form
 \begin{equation}
\langle O^{(s)(jj')}(x)\rangle_{\mathcal{C}_A^{(n)}}=C_{(jj')}^A\langle\phi^{(j)}(x)\phi^{(j)}(x_A)\rangle_M\langle\phi^{(j')}(x)\phi^{(j')}(x_A)\rangle_M+\cdots\,,
\end{equation}
Note that in the transverse direction, we still have the identification $\theta \sim \theta +2\pi n$. Consequently we can apply the $1/n$ prescription in the bulk computation as well. In other words, in the $n \to 1$ limit, the fields can always be taken as the free fields, and the possible interaction can be ignored safely.

Before doing bulk calculations in details, let us first explain our conventions. In the following we use the capital alphabets $M,N,\cdots$ to denote the bulk indices, taking values from $0$ to $d$. The bulk coordinates are denoted by $r^M=(t_E\,,x^i,\, z)$ where $i=1,\dots,(d-1)$. We refer $r^0$ to the Euclidean time  $t_E$ and  $r^d$ to the radial coordinate $z$. We  work in the Poinc\'{a}re coordinates for the bulk metric and set the AdS radius to unity. Since now $z$ denotes the radial coordinate in the bulk, the cross ratio will be denoted by $z_{cr}$. 

 For any two points $r=(t_E,x^i,z),r'=(t'_E,x'^i,z')$ in the $AdS_D$ ($D={d+1}$) vacuum, one can always connect them by a geodesic whose length is
\begin{equation}
 \ell(r\,,r')=\log{\Big( \fft{1+\sqrt{1-\xi^2}}{\xi} \Big)}\,,
 \end{equation}
where
\begin{equation}
\xi(r\,,r')=\fft{2zz'}{z^2+z'^2+(t_E-t'_E)^2+(\vec{x}-\vec{x}')^2}\,,
\end{equation}
 is a biscalar.
In many cases, it is convenient to introduce the chordal distance $u(r\,,r')$
\begin{equation}
 u\equiv \cosh{\ell}-1=\xi^{-1}-1 \,.
\end{equation}
We denote the bulk covariant derivative as $D_{M}$ and
\begin{equation}
 \partial_M\equiv\fft{\partial}{\partial r^M}\,,\quad  \partial_{M'}\equiv\fft{\partial}{\partial {r'}^{M}}\,.
 \end{equation}
 Since the distance between the two hemispheres are much larger than their radius, we have
 \be \xi\simeq \fft{2zz'}{|x-x'|^2}\rightarrow 0\,, \ee
 and hence
 \be
 u\simeq 1/\xi\rightarrow \infty.
 \ee
  This is a useful relation throughout this section.

\subsection{Scalar field}
As a warm up, let us first calculate the leading order MI from a free scalar field reported in \cite{AF2015}. For a free scalar with mass square $m^2=\Delta(\Delta-d)$, it is dual to a scalar primary operator with dimension $\Delta$ on the boundary CFT. Its bulk-to-bulk propagator is
\begin{equation}\label{scalarpropagator}
\langle \phi(r)\phi(r') \rangle=C_\Delta \Big(\fft{\xi}{2} \Big)^\Delta F\Big(\fft{\Delta}{2}\,,\fft{\Delta+1}{2}\,,\nu+1;\xi^2\Big) \,,
\end{equation}
where $\nu=\sqrt{m^2+d^2/4}$ and $C_\Delta$ is a normalization constant.
For the reference points $r_A\,,r_B$, in the large distance limit, we have $z_A \,,z_B\ll |x_A-x_B| $. Then
\begin{equation}
 \langle \phi(r_A)\phi(r_B) \rangle\simeq C_\Delta \fft{z^\Delta_A z^\Delta_B}{|x_A-x_B|^{2\Delta}} \,.
 \end{equation}
Thus, we find that
\begin{equation}
 I(A,B)|_{s=0}=\fft{1}{|x_A-x_B|^{4\Delta}}\lim_{n\rightarrow 1} \fft{1}{2(n-1)}\sum_{j\neq j'}\widetilde{C}^A_{(jj')}\widetilde{C}^B_{(jj')} \,,\label{MIscalar}
 \end{equation}
where
$
\widetilde{C}^A_{(jj')}\equiv C_\Delta z^{2\Delta}_A C^A_{(jj')}.$
 To calculate the OPE coefficients, we do a coordinate transformation similar to (\ref{ct})
\begin{equation}
r'^M=\fft{r^M+ n^M R_A}{\Omega}-\fft{n^M}{2R_A} \,,\label{bct}
\end{equation}
where $n^M=(0,1,0,\dots,0)$ is a $D$-dimensional unit vector, and
\begin{equation}
\Omega=R^2_A+2R_A\, x_1+z^2+t_E^2+\vec{x}^2\,.
\end{equation}
This transformation  preserves the AdS metric. It should be emphasized that this is not a conformal transformation any longer.  Under this transformation the original conifold geometry $\mathcal{C}_A^{(n)}$ with spherical conical singularity ($t_{\mathrm{E}}=0\,,z^2+\delta_{ij}x^i x^j=R_A^2$) is mapped to a new
conifold geometry $\mathcal{C}_A^{'(n)}$ with the singularity located at the plane ($t'_{\mathrm{E}}=0\,,x'^1=0$). The infinity is mapped to a finite point
\begin{equation}\label{rfp}
r_{\infty}=(z_\infty\,,t_{\mathrm{E}}=0\,,x^i_\infty)\hs{2ex}\longrightarrow \hs{2ex}r'_{\infty}=(z'=\epsilon\,,t'_{\mathrm{E}}=0\,,x'^i=-\fft{n^i}{2R_A})  \,,
\end{equation}
where the large separation limit corresponds to $\epsilon\rightarrow 0$. To further simplify our calculations, we take the reference point to be $r_A=(z_A\,,t_E=0\,,x^i=0)$ which is mapped to
\begin{equation}\label{rap}
 r_A=(z_A\,,t_E=0\,,x^i=0) \hs{2ex}\longrightarrow \hs{2ex}r'_A=\Big(z'_A=\fft{z_A}{R^2_A+z^2_A}\,,t'_E=0\,,x'^i=\fft{n^i}{2R_A}\fft{R^2_A-z^2_A}{R^2_A+z^2_A} \Big) \,.
 \end{equation}
For $r_\infty$, the $\Omega$ factor $\Omega(r_\infty)\simeq x_\infty^2$ and for $r_A$, $\Omega(r_A)=R_A^2+z_A^2$.

Under the coordinate transformation (\ref{bct}), at the leading order a bulk spin-$s$ operator transforms as
\begin{equation}\label{eq:trans}
 \langle O_{M_1 M_2\dots M_s}(r_\infty)\rangle_{\mathcal{C}_A^{(n)}}\approx \Omega^{-s}(r_\infty)I^{N_1}_{M_1}(r_\infty)I^{N_2}_{M_2}(r_\infty)\dots I^{N_s}_{M_s}(r_\infty)\langle O_{N_1 N_2\dots N_s}(r'_\infty)\rangle_{\mathcal{C}_A^{'(n)}} \,.
 \end{equation}
In fact, due to the rotational symmetry, we only need to consider the time-time-...-time component \cite{CCHL2017}
\begin{eqnarray}\label{bulkspin}
&&\langle O_{00\dots 0}(r_\infty)\rangle\simeq x_\infty^{-2s}\,\langle O_{00\dots 0}(r'_\infty)\rangle\,,\nn\\
&&\langle O_{00\dots 0}(r_A)\rangle= (R_A^2+z_A^2)^{-s}\,\langle O_{00\dots 0}(r'_A)\rangle\,.
\end{eqnarray}
With all these results in hand, we are ready to compute the OPE coefficients. For the scalar field, we find
\begin{equation}
 C_{(jj')}^A =\Big(C_\Delta\epsilon^{\Delta}z_A^{\Delta}\Big)^{-2}\langle O^{(s)(jj')}(r'_\infty)\rangle_{\mathcal{C}_A^{'(n)}}\,.
\end{equation}
Here the one-pint function on the right-hand side can be computed using the $1/n$ prescription
\begin{equation}
\left\langle O^{(s)(jj')}(r'_{\infty})\right\rangle _{\mathcal{C}'{}_{A}^{(n)}}=\langle \phi^{(j)}(r'_\infty)\phi^{(j')}(r'_\infty)\rangle_{\mathcal{C}_A^{'(n)}}= \left(\frac{2C_{\Delta}}{\nu}\right)\epsilon^{2\Delta}R_{A}^{2\Delta}\sin^{-2\Delta}\frac{\theta_{jj'}}{2n}\,,
\end{equation}
where $\theta_{jj'}\equiv\theta_j-\theta_{j'}$. So we get
\begin{equation}
C_{(jj')}^{A}=\frac{\nu R_{A}^{2\Delta}}{2C_{\Delta}(z_{A})^{2\Delta}}\sin^{-2\Delta}\frac{\theta_{jj'}}{2n}\,.
\end{equation}
Substituting the above results into (\ref{MIscalar}), we finally obtain
\begin{equation}
I(A,B)|_{s=0} =  \frac{\sqrt{\pi}}{4^{2\Delta+1}}\frac{\Gamma(2\Delta+1)}{\Gamma(2\Delta+\frac{3}{2})}z_{cr}^{2\Delta}
\end{equation}
in which $z_{cr}=\frac{4R_{A}R_{B}}{|x_{A}-x_{B}|{}^{2}}$. This is exactly matched with the boundary result (\ref{coeffscalar}) of a primary scalar operator with the scaling dimension $\Delta$, as we expected. %We can get the same result using the image method, as had been done in \cite{AF2015}.

We continue to construct a spin-1 operator from the bulk scalar fields residing at different replicas. We propose that the vector operator
\begin{equation}
 O^{(s)(jj')}_M\equiv \phi^{(j)}\partial_M \phi^{(j')}-(j\leftrightarrow j')\,,
 \end{equation}
is dual to the spin-1 operator (\ref{Jscalar}) with the scaling dimension $2\Delta+1$ in the boundary  theory.
A straightforward calculation shows that its time-time component of the propagator in the large distance limit is
\begin{equation}
 \langle O^{(s)(jj')}_0(r_A)O^{(s)(jj')}_{0}(r_B) \rangle\simeq 4\Delta \fft{C^2_\Delta z_A^{2\Delta}z_B^{2\Delta} }{|x_A-x_B|^{2(2\Delta+1)}}\,,
 \end{equation}
and
\begin{equation}
 \langle O^{(s)(jj')}_0(r'_\infty)O^{(s)(jj')}_{0}(r'_A) \rangle=4\Delta C^2_\Delta \epsilon^{2\Delta} (R^2_A+z_A^2) \,.
 \end{equation}
Using the coordinate transformation (\ref{eq:trans}) and the $1/n$ prescription, we read
\begin{equation}
C_{(0j)}^{(A)0}=\frac{\nu R_{A}^{2\Delta+1}}{2C_{\Delta}z_{A}^{2\Delta}}\frac{1}{n}\cos\frac{\theta_{j}}{2n}\sin^{-2\Delta-1}\frac{\theta_{j}}{2n}.
\end{equation}
It follows that at the leading order the mutual information from the spin-1 field is given by
\begin{equation}
I(A,B)|_{s=1}= -\frac{\Delta}{2}\frac{\sqrt{\pi}}{4^{2\Delta+1}}\frac{\Gamma(1+2\Delta)}{\Gamma(\frac{5}{2}+2\Delta)}z_{cr}^{2\Delta+1}.
\end{equation}
This exactly matches with the boundary result (\ref{coeffJs}). Note that it is negative.

Next we construct a bulk spin-2 operator from the original scalar field as
\begin{equation}\label{bulkspin2}
O^{(s)(jj')}_{M N}\equiv \fft 12 P^{E F}_{M N}\Big( D_E \phi^{(j)}D_F \phi^{(j')}-\fft{\Delta}{\Delta+1} \phi^{(j)}D_E D_F \phi^{(j')}+(j\leftrightarrow j') \Big) \,,
\end{equation}
where the bulk projector is defined to be
\begin{equation}\label{P2}
 P_{M N}^{E F}\equiv h^{E}_{(M}h^{F}_{N)}-\fft 1d\, h_{M N}h^{E F}\,,\quad h_{M N}=g_{M N}-n_M n_N.
  \end{equation}
Here $n^M=(0\,,\sqrt{g^{zz}}\,,\cdots\,,0)$ is the unit normal vector of the time-like hypersurface orthogonal to the radial direction in the Poinc\'{a}re coordinates. $h_{M N}$ is the induced metric on the constant $z$ hypersurface. Note that the projector is symmetric and traceless. The above spin-2 field is dual to the boundary spin-2 operator (\ref{tscalar}) with the scaling dimension $2(\Delta+1)$. However, there is something subtle in the above definition that should be clarified. To show this, we recall the definition of the boundary spin-2 operator\cite{CCHL2017}
\be O^{(s)(jj')}_{\mu\nu}=\fft 12 \mathcal{P}^{\alpha\beta}_{\mu\nu}\Big(\partial_\alpha \mathcal{O}^{(j)}\partial_\beta \mathcal{O}^{(j')}-\fft{\Delta}{\Delta+1}\mathcal{O}^{(j)}\partial_\alpha\partial_\beta \mathcal{O}^{(j')}+(j\leftrightarrow j')  \Big) \,,\ee
where the boundary projector is defined in (\ref{tscalar}) using the Euclidean metric.
Naively, one may expect that the gravity duals to the higher spin operators in the boundary CFT can be constructed via a minimally replacing rule. That is by replacing $\mathcal{O}\rightarrow \phi\,,\partial_\mu\rightarrow \nabla_M\,,\delta_{\mu\nu}\rightarrow g_{MN}$ in the boundary operators, one obtains the dual bulk fields. In this case, the bulk spin-2 projector would be
\be \widetilde{P}_{MN}^{EF}=g^{E}_{(M}g^{F}_{N)}-\fft 1 D\, g_{M N}g^{E F}\,,\ee
where a tilde is used to distinguish it from the projector (\ref{P2}). However, this is not correct and cannot produce the correct answers. The correct bulk projector should be (\ref{P2}).  It looks unnatural at first glance since the projector is defined on a time-like hypersurface instead of the AdS bulk. Nonetheless, we have a simple interpretation how it works. The projector plays two-fold roles. Firstly, it maps a bulk operator onto the time-like hypersurface $z=\mathrm{const}$, suppressing all the radial components. Secondly, the operators on the hypersurface are projected to be symmetric and traceless. In this sense, the bulk spin-2 field defined in (\ref{bulkspin2}) can be viewed as living on the curved $d$-dimensional sub-manifold with $z=\mathrm{const}$, which can be obtained by extending the boundary (and its field theory content) into the deep bulk region. On the other hand, when close to the boundary, one finds $\phi\rightarrow z^\Delta \mathcal{O}\,,h_{MN}\rightarrow z^{-2}\delta_{\mu\nu}$ such that $P_{MN}^{EF}\rightarrow \mathcal{P}_{\mu\nu}^{\alpha\beta}$ and $O^{(s)(jj')}_{MN}\rightarrow z^{2\Delta}O^{(s)(jj')}_{\mu\nu}$.
 Here note that the prefactor in the spin-2 operator is $z^{2\Delta}$ instead of $z^{2\Delta+2}$. It should be so because the bulk spin-2 field has a scaling dimension $2\Delta$. Now it becomes clear that our bulk spin-2 field defined in (\ref{bulkspin2}) is indeed dual to the spin-2 operator in the boundary CFT. Following our discussions, it is easy to construct the gravity duals for general higher spin operators in the boundary theory that are carefully studied in \cite{CCHL2017}.

The remaining calculations are straightforward. At the large separation limit, the relevant two-point function is
\begin{equation}
  \langle O^{(s)(jj')}_{00}(r_A) O^{(s)(jj')}_{00}(r_B) \rangle\simeq\fft{4(d-1)\Delta^2(2\Delta+1)}{d(\Delta+1)}\fft{C^2_\Delta z_A^{2\Delta}z^{2\Delta}_B}{|x_A-x_B|^{4(\Delta+1)}}\,,
 \end{equation}
and
\begin{equation}
  \langle O^{(s)(jj')}_{00}(r'_\infty) O^{(s)(jj')}_{00}(r'_A) \rangle=C^2_\Delta\, \epsilon^{2\Delta} z_A^{2\Delta}(R_A^2+z^2_A)^2\,\fft{4(d-1)\Delta^2(2\Delta+1)}{d(\Delta+1)}\,.
 \end{equation}
The corresponding OPE coefficients are given by
\begin{equation}\label{eq:C00jj}
 C^{(A)00}_{(jj')}=(R_A^2+z_A^2)^2\fft{\langle O^{(s)(jj')}_{00}(r'_\infty)
 \rangle_{C^{(n)}_{A'}}}{\langle O^{(s)(jj')}_{00}(r'_\infty) O^{(s)(jj')}_{00}(r'_A) \rangle}\,.
 \end{equation}
According to the $1/n$ prescription, we find
\begin{align}
C_{(0j)}^{(A)00}=-\frac{\nu z_{A}^{2\Delta}R_{A}^{2\Delta+2}}{4C_{\Delta}\Delta(1+2\Delta)}\frac{(1+n^{2}+2\Delta)}{n^{2}}\left((1+2\Delta)\sin^{-2\Delta-2}\frac{\theta_{j}}{2n}-2\Delta\sin^{-2\Delta}\frac{\theta_{j}}{2n}\right).
\end{align}
Note that we must work in the $(r,\theta)$ coordinate system to derive the one-point function  in the replicated geometry.
After some simple calculations, we finally get
\begin{equation}
I(A,B)|_{s=2}=\frac{(d-1)}{d}\frac{2+4\Delta+3\Delta^{2}}{(1+2\Delta)^{2}}\frac{2\sqrt{\pi}}{4^{2\Delta+3}}\frac{\Gamma(2\Delta+3)}{\Gamma(2\Delta+\frac{7}{2})}z_{cr}^{2\Delta+2}.
\end{equation}
It  matches exactly with the boundary result (\ref{b2dp22}) of the spin-2 operator.

\subsection{Gauge Bosons}

Now we generalize the discussions to the  massless gauge fields in the bulk, which are dual to the conserved currents in the boundary. For the vector-type operators in the boundary CFT, the bulk dual should be vector fields. In general, the vector field is massive and there is no gauge symmetry. In this subsection, we focus on the case that the bulk field is a gauge field, and leave the discussion on the massive case to Appendix A. The gauge field is interesting as it often appears in the spectrum of AdS supergravity. Moreover, the computation of the mutual information due to the exchange of the gauge field presents novel feature, which we would like to report.

For a U(1) gauge boson, the bulk-to-bulk propagator is given by\cite{Gauge1998}
\begin{eqnarray}\label{gauge boson}
G_{M N}(r,R)&=&-\partial_M\partial_{N'} u F(u)+\partial_ M \partial_ {N'} S(u)\nonumber\\
&=&\partial_ M u\partial_{N'} u S''(u)+\partial_ M \partial_{N'} u (S'(u)-F(u)),
\end{eqnarray}
where the function $F(u)$ is
\be F(u)=\frac{\Gamma \left(\frac{d-1}{2}\right)}{4\pi ^{(d+1/)2}}\frac{1}{\big(u(u+2)\big)^{(d-1)/2}},\ee
and $S(u)$ is a gauge artifact. In the Feynman gauge, it is determined by \cite{Gauge1998}
\be u(u+2)S'''+(d+3)(u+1)S''+(d+1)S'=2F.\ee
The explicit expression for $S(u)$ is complicated, but we only need its asymptotic behavior since we are considering the MI between two far separated regions. In the large separation limit, we have
\begin{eqnarray}
F(u)&\simeq&\frac{\Gamma \left(\frac{d-1}{2}\right)}{4\pi ^{(d+1/)2}}\frac{1}{u^{d-1}},\hs{3ex}
S'(u)\simeq \frac{\Gamma \left(\frac{d-1}{2}\right)}{4\pi ^{(d+1/)2}}\frac{1}{(2-d) u^{d-1}}.
\end{eqnarray}

In general, the gauge part gives vanishing contributions when integrated against conserved currents, for example in  Witten diagrams considered in \cite{Gauge1998}. This is because the surface terms in the partial integration vanish. However, in our case, the situation is quite different owing to the presence of additional boundaries:  the two separated entangling surfaces. We find that the gauge part $S(u)$ of the propagator, besides the usually called physical part $F(u)$, also contributes to the leading order of the MI. This seems in conflict with what we cognize before since clearly the MI should be gauge independent. To clarify this, let us first present our results in details.

To compare with the boundary results, we construct two kinds of operators with spin 0 and spin 2 respectively
\begin{align}\label{O0O1}
O^{(v)(jj')}= & h^{MN}A_{M}^{(j)}A_{N}^{(j')},\hs{3ex}
O_{AB}^{(v)(jj')}=  P_{AB}^{MN}A_{M}^{(j)}A_{N}^{(j')},
\end{align}
where the bulk projector $P_{AB}^{MN}$ is defined in (\ref{P2}) and superscript $(v)$ stands for operators constructed from the vector gauge boson. As will be shown shortly, the above two operators are dual to the boundary operators with the same spins constructed from a current operator with the scaling dimension $\Delta=d-1$.

For the spin-0 operator, the calculation is similar to the discussions for the scalar operators except that we now need a different propagator. Using the propagator for the gauge boson (\ref{gauge boson}), we find at the leading order in the large distance limit
\begin{align}
\left\langle O^{(v)(jj')}(r_{A})O^{(v)(jj')}(r_{B})\right\rangle _{M}\simeq \frac{\Gamma \left(\frac{d-1}{2}\right)^2}{16 \pi ^{d+1}}\frac{d (d-1)^2}{(d-2)^2}\frac{\left(2 z_A z_B\right){}^{2 (d-1)}}{|x_A-x_B|^{4 (d-1)}}.
\end{align}
It is worth emphasizing that this result is derived from the total bulk-to-bulk propagator of the gauge boson including the gauge part.  Applying the $1/n$ prescription, we  get the OPE coefficient
\begin{equation}
C_{(0j)}^{A}=\frac{4\pi^{(1+d)/2}(d-2)}{\Gamma\left(\frac{d-1}{2}\right)d(d-1)2^{d-1}}z_{A}^{2-2d}R_{A}^{2d-2}\frac{(d-1)n^{2}-1}{n^{2}}\sin^{2-2d}\left(\frac{\theta_{j}}{2n}\right).
\end{equation}
The bulk MI turns out to be
\begin{equation}\label{gaugespin0}
I(A, B)|_{s=0}=\frac{(d-2)^2}{4^{2 d-1}d}\frac{\sqrt{\pi } \Gamma (2 d-1)}{\Gamma \left(2 d-\frac{1}{2}\right)}z_{cr}^{2(d-1)}.
\end{equation}

For the spin-2 operator, its contribution to the MI at the leading order is
\begin{equation}
I(A,B)|_{s=2}=\lim_{n\to1}\frac{n}{2 (n-1)}\sum _{j=0}^{n-1} C_{(0j)}^{(A)KL} C_{(0 j)}^{(B)MN} \left\langle O_{MN}^{(v)(0j)}\left(r_A\right) O_{KL}^{(v)(0j)}\left(r_B\right)\right\rangle.
\end{equation}
As discussed before, for the higher spin operators only the time-time component has non-trivial contributions to the MI at the leading order. Hence, without loss of generality, we can drop the other components in the following. At the large separation limit, we find
\begin{equation}
\left\langle O_{00}^{(v)(jj')}\left(r_A\right) O_{00}^{(v)(jj')}\left(r_B\right)\right\rangle \simeq\frac{\Gamma^2(\frac{d-1}{2})}{4\pi ^{d+1}}\frac{(d-1)^3}{d(d-2)^2}\frac{(2z_Az_B)^{2d-4}}{|x_A-x_B|^{4d-4}}.
\end{equation}
Both the physical part $F(u)$ and the gauge part $S(u)$ in the propagator contribute. The OPE coefficient  is
\begin{equation}
C_{(0j)}^{(A)00}= \frac{\pi^{(1+d)/2}2^{3-d}(2-d)}{(d-1)\Gamma\left(\frac{d-1}{2}\right)}z_{A}^{4-2d}R_{A}^{2d-2}\frac{(1+n^{2})}{n^{2}}\sin^{2-2d}\left(\frac{\theta_{j}}{2n}\right).
\end{equation}
Then taking the limit $n\to 1$, we finally get
\begin{equation}\label{gaugespin2}
I(A, B)|_{s=2}=  \frac{d-1}{d}\frac{\sqrt{\pi}}{4^{2d-2}}\frac{\Gamma(2d-1)}{\Gamma(2d-\frac{1}{2})}z_{cr}^{2(d-1)}.
\end{equation}
Now we are able to compare our bulk results with the boundary results reported in \cite{CCHL2017}. We find that our results are perfectly matched with (\ref{vector0}, \ref{vector2}) of a current operator which has scaling dimension $\Delta=d-1$. Indeed, according to the AdS/CFT correspondence, a gauge boson in the bulk is dual to a conserved current in the boundary CFT. In this sense, the above results may be expected at the very start. However, the subtlety is that throughout our calculations, the gauge part of the bulk-to-bulk correlator associated with the function $S(u)$  in the two-point functions has non-trivial contribution to the holographic MI for both the spin-0 and  spin-2 operators. It is remarkable that from our discussions, the usually so-called gauge artifact $S(u)$ always has significant contributions to the bulk mutual information. But this does not mean the result is gauge dependent. In fact,  we find that, to the leading order in $u$, the contribution from the gauge part is independent of the gauge parameter.

One possible interpretation for the nonvanishing contribution from the gauge part could be that the gauge symmetry is effectively broken around the entangling surfaces, giving rise to an extra physical degree of freedom living on the boundaries.  In fact, from the leading order MI  for a massive vector field in the bulk (the calculation details is given in Appendix A.1), we find that the results are exactly matched with boundary results of a current operator with a generic scaling dimension $\Delta$ as well. This supports our argument since in general a massive vector field has one more degree of freedom than the gauge boson but they both give the same results to the MI at the leading order.

An interesting question is how to understand  the breaking of gauge symmetry on the entangling surfaces. We recall the definition of the entanglement entropy (\ref{eq:ee}) for any subsystem $A$. The key ingredient is the reduced density matrix of $A$ which is defined by tracing out the degrees of freedom in its complement $\bar A$ from the total density matrix. However, for gauge theories the  Hilbert space of physical states can not be factorized into a tensor product of the Hilbert space of the states localized in the spatial regions $A$ and $\bar A$. In \cite{Buividovich:2008gq}, it was argued that the elementary excitations in gauge theories are electric strings, which are closed loops rather than points in space. Hence, it is indispensable that there are closed loops which are belong to both $A$ and $\bar A$. So the reduced density matrix of $A$ can only be well defined if the Hilbert space of physical states is extended by including the states of electric strings that open on the boundary of $A$. The endpoints of the electric strings on the boundary were previously pure gauge degrees of freedom but now become physical and hence break gauge symmetries, giving rise to extra contributions to the entanglement entropy. We refer the readers to literatures such as \cite{Donnelly:2011hn,Casini:2013rba,Donnelly:2014gva,Donnelly:2016auv} for more discussions on this issue. It will be of great interests to compute the quantum corrections of gauge fields to the entanglement entropy for a single entangling surface in the AdS/CFT correspondence. We leave this as a direction for future research.

%When doing this, one needs impose proper boundary conditions on the entangling surface $\partial A$. This is even clearer when we think about it in the path-integral representation. Probably, the presence of additional boundaries leads to non-vanishing surface terms (on the entangling surfaces) associated with the gauge artifact $S(u)$ in partial integrations. This gives rise to extra contributions to the entanglement entropy in contrast to the usual cases where the boundary is placed at infinity. It deserves further investigations whether such  calculations can be performed analytically and matched with our results using the OPE technique.

%To end this subsection, we argue that our bulk results at leading order for gauge boson can be interpreted as the entanglement of the new emergent degrees of freedom living on the two separated half-spheres plus the contributions from the transverse modes of the gauge boson.

\subsection{Gravitons}

Now we consider the contribution of the massless graviton denoted by $\tilde{G}_{MN}$. This is dual to the conserved stress tensor in the boundary theory. For a general spin-2 operator in the boundary, the corresponding field could be a massive graviton, whose leading order contribution to the mutual information is put in Appendix A.2. Here we focus on the massless case.

The bulk-to-bulk propagator of the graviton can be written as
\begin{equation}\label{eq:gra}
\langle \tilde{G}_{MN} (x_A) \tilde{G}_{E'F'}(x_B)\rangle = G_{MN,E'F'}(x_{A},x_{B})=g_{MN}g_{E'F'}T+\sum_{i=1}^{5}G^{(i)}O_{MN,E'F'}^{(i)},
\end{equation}
in which the explicit forms of coefficients $T,G^{(i)}$ and the tensor structures $O_{MN,E'F'}^{(i)}$ in the Landau gauge can be found in \cite{Graviton1999_1,Graviton1999_2}. For self-consistency, we list them in Appendix B. Actually, we only need the asymptotic behavior of the propagator in the large distance limit. Note that the physical part of the propagator has the form
\begin{equation}
G_{MN,EF}=(\partial_{M}\partial_{E'}u\partial_{N}\partial_{F'}u+\partial_{M}\partial_{F'}u\partial_{N}\partial_{E'}u)\tilde{G}(u)+g_{MN}g_{E'F'}\tilde{H}(u).
\end{equation}
The relations between coefficients $\tilde{G}(u),\tilde{H}(u)$ and $T,G^{(i)}$ can be found in \cite{Graviton1999_1}. Note that similar to the gauge boson case, the gauge part of the propagator gives significant contributions to the mutual information such that the bulk result agrees with the boundary result. % The gauge part in the propagator of graviton contributes to the holographic MI just like in the gauge boson case.

There are three kinds of bulk operators which have the leading contributions to the mutual information
\begin{align}
&O^{(t)(jj')}=  h^{MN}h^{EF}\tilde{G}_{MN}^{(j)}\tilde{G}_{EF}^{(j')}\label{O21}\,,\\
&O_{IJ}^{(t)(jj')}=  P_{IJ}^{AB}h^{CD}\tilde{G}_{AC}^{(j)}\tilde{G}_{BD}^{(j')}\label{O22}\,,\\
&{O}_{ABCD}^{(t)(jj')}=  P_{ABCD}^{EFGK}\tilde{G}_{EG}^{(j)}\tilde{G}_{FK}^{(j')}\label{O4}\,,
\end{align}
in which superscript $(t)$ stands for the operators constructed from the gravitons.  $P_{IJ}^{AB}$ is defined by (\ref{P2}) and
\begin{eqnarray}
P_{ABCD}^{EFGK}&=&\frac{1}{24}h^{E}_{(A}h^{F}_{B}h^{G}_{C}h^{K}_{D)}-\frac{h_{(AB}h^{(EF}h^G_Ch^{K)}_{D)}}{12(d+4)}
+\frac{h_{(AB}h_{CD)}h^{(EF}h^{GK)}}{3(d+2)(d+4)}.
\end{eqnarray}

For the spin-0 and the spin-2 cases, the calculations are similar to the discussions for the U(1) gauge boson. In the large separation limit, the two-point function of the spin-0 operator is
\begin{equation}
\left\langle O^{(t)(jj')}(r_{A})O^{(t)(jj')}(r_{B})\right\rangle _{M}\simeq a_0(D)\frac{(2z_{A}z_{B})^{2d}}{|x_{A}-x_{B}|{}^{4d}}\,,
\end{equation}
 in which
\begin{align}
a_{0}(D)=\frac{2^{2d+3}d^{2}}{(d+1)^{2}(d-1)(d+2)}b(D)^{2}.
\end{align}
Here the factor $b(D)$ is given in (\ref{eq:bn}) in Appendix B and $D=d+1$. Having the two-point function and using the $1/n$ prescription, we deduce the corresponding OPE coefficient
\begin{equation}
C_{(0j)}^{A}=\frac{4b(D)d\left(1-(d-1)n^{2}+\frac{(d+1)(d-2)}{2}n^{4}\right)}{a_{0}(D)(d+2)(d+1)(d-1)n^{4}}z_{A}^{-2d}R_{A}^{2d}\sin^{-2d}\frac{\theta_{j}}{2n}.
\end{equation}
It follows that the leading holographic MI is
\begin{equation}
I(A, B)\big|_{s=0}=\frac{1}{2^{4 d+3}}\frac{(d-1) (d-2)^2}{d+2}\frac{\sqrt{\pi } \Gamma (2 d+1)}{\Gamma \left(2 d+\frac{3}{2}\right)}z_{cr}^{2d},
\end{equation}
which is in exact agreement with (\ref{tensor0}) provided $\Delta=d$.

For the spin-2 operator, the time-...-time component of the two-point function is
\begin{align}
\left\langle O_{00}^{(t)(jj')}(r_{A})O_{00}^{(t)(jj')}(r_{B})\right\rangle _{M}= & a_2(D)\frac{(2z_{A}z_{B})^{2d-2}}{|x_{A}-x_{B}|{}^{4d}}\,,
\end{align}
where
 \be
 a_{2}(D)= 2^{2d+4}\frac{(d-2)(d+4)}{(d-1)(d+1)^{2}(d+2)^{2}}b(D)^{2}.
 \ee
After some calculations we get the OPE coefficient
\begin{equation}
C_{(0j)}^{(A)00}=\frac{b(D)}{a_{2}(D)}\frac{8\left(-2+(d-2)n^{2}+dn^{4}\right)}{(d+2)(d+1)n^{4}}z_{A}^{-2d+2}R_{A}^{2d}\sin^{-2d}\frac{\theta_{j}}{2n},
\end{equation}
and the contribution of the spin-2 operator to the holographic MI
\begin{equation}
I(A, B)\big|_{s=2}=\frac{(d-2) (d-1)}{2^{4 d} (d+4)}\frac{\sqrt{\pi } \Gamma (2 d+1)}{\Gamma \left(2 d+\frac{3}{2}\right)}z_{cr}^{2d},
\end{equation}
which is in agreement with (\ref{tensor2}) when $\Delta=d$.

For the spin-4 operator (\ref{O4}), its contribution to the MI can be formally written as
\begin{equation}
I(A, B)|_{s=4}=\lim_{n\to 1}\frac{n}{2 (n-1)}\sum _{j=1}^{n-1} C^{(A)ABCD}_{(0j)} C^{(B)EFHI}_{(0j)} \left\langle O_{ABCD}^{(t)(0j)}\left(r_A\right) O_{EFHI}^{(t)(0j)}\left(r_B\right)\right\rangle\,,
\end{equation}
where $ C^{(A)ABCD}_{(0j)}$ is the corresponding expansion coefficient for the subregion $A$.
As emphasized earlier, only the time-...-time component is relevant for our purpose, namely
\begin{equation}\label{gravitonmi}
I(A, B)|_{s=4}=\lim_{n\to 1}\frac{n}{2 (n-1)}\sum _{j=1}^{n-1} C^{(A)0000}_{(0j)} C^{(B)0000}_{(0j)} \left\langle O_{0000}^{(t)(0j)}\left(r_A\right) O_{0000}^{(t)(0j)}\left(r_B\right)\right\rangle.
\end{equation}
At the large separation, the two-point function of the operator at the leading order is given by
\begin{equation}
\left\langle O_{0000}^{(t)(jj')}(x_{A})O_{0000}^{(t)(jj')}(x_{B})\right\rangle\simeq a_4(D)\frac{(2z_{A}z_{B})^{2d-4}}{|x_{A}-x_{B}|{}^{4d}}\,,
\end{equation}
where
\begin{equation}
a_{4}(D)=\frac{4^{d+4}d^{2}}{(d-1)(d+1)(d+2)^{3}(d+4)}b(D)^{2}.
\end{equation}
The coefficient $ C^{(A)0000}_{(0j)} $ can be read by using the $1/n$ prescription
\begin{eqnarray}
C_{(0j)}^{(A)0000}=-\frac{b(D)}{a_{4}(D)}\frac{2^{6}d(1+n^{2})^{2}}{(d+4)(d+2)^{2}n^{4}}z_{A}^{-2d+4}R_{A}^{2d}\sin^{-2d}\frac{\theta_{j}}{2n}.
\end{eqnarray}
Finally, plugging the above results into (\ref{gravitonmi}), we obtain
\begin{equation}
I(A, B)|_{s=4}=\frac{(d+1) (d-1)}{4^{2 d-1} (d+4) (d+2)}\frac{\sqrt{\pi } \Gamma (2 d+1)}{\Gamma \left(2 d+\frac{3}{2}\right)}z_{cr}^{2d}\,,
\end{equation}
which is in good match with (\ref{tensor4}) when $\Delta=d$.

All these  contributions of the graviton to the  holographic MI are perfectly matched with those of a spin-2 primary operator which has scaling dimension $\Delta=d$ in the boundary CFT. However, to compare the results with the stress tensor in the CFT, we need to clarify some subtleties in the CFT side. % since the stress tensor is quasi-primary rather than primary.

First, although interchanging a single operator $T_{\mu\nu}^{(j)}$ does not contribute to the mutual information, the derivative of the R\'enyi mutual information with respect to $n$ in the $n\rightarrow 1$ limit contains some universal information about the underlying theories as well. To be precise, we have
\be\label{renyi} I_n(A\,,B)=\fft{d(d-1)}{4\pi^2 C_T}\,\fft{h_n^2}{n(n-1)}z_{cr}^d+\dots \,,\ee
where $h_n$ is the conformal dimension of the higher dimensional twist operators. For convenience, we introduce its definition from the long-distance behavior of the stress tensor, namely $|x|\rightarrow \infty$,
\be \langle T_{00}(0,x) \rangle_{\mathcal{C}_A^{(n)}}=\fft{d-1}{2\pi}\Big(\fft{2R_A}{|x|^2}\Big)^d h_n\,,\quad \langle T_{ij}(0,x) \rangle_{\mathcal{C}_A^{(n)}}=-\fft{\delta_{ij}}{2\pi}\Big(\fft{2R_A}{|x|^2}\Big)^d h_n \,,\ee
and all the other components are zero at $t_E=0$. Furthermore, it was proved in \cite{Hung:2014npa} that although the conformal dimension $h_n$ vanishes in the limit $n\rightarrow 1$, its derivative with respect to $n$ gives a non-trivial universal result
\be \partial_n h_n|_{n=1}=2\pi^{(d+2)/2}\fft{\Gamma{(d/2)}}{\Gamma{(d+2)}}\,C_T \,.\ee
Consequently, the mutual information does not receive contributions from exchange of a single operator but its derivative does. We find
\be \partial_n I_n|_{n=1}=d(d-1)\pi^d \fft{\Gamma^2{(d/2)}}{\Gamma^2{(d+2)}}\,C_T\,z_{cr}^d+\cdots \,.\ee
This result is consistent with (5.16-5.18) in \cite{Hung:2014npa}, in which the correlator of spherical twist operators around $n=1$ was derived in a different approach. It is easily seen that the derivative of the mutual information at the order $z_{cr}^d$ contains universal information about the underlying theories,
which is however not seen at the leading order $z_{cr}^{2d}$ result in the mutual information itself. This is interesting and probably could be generalized to generic higher spin operators. %For the current operator the authors in \cite{Belin:2013uta} also arrived at similar results for the coefficient of the one-point function defined in the conifold geometries. Hence, we expect that the derivative of the mutual information for the exchange of a single current operator contains non-trivial universal information at the order $z_{cr}^\Delta$ about the CFT.

Second, as the stress tensor is a quasi-primary operator, it is known that anomalous terms should be included when it transforms under a conformal transformation. One may worry about that the calculations for the stress tensor will become much more complicated than a primary spin-2 operator. Fortunately, we find that its connected two-point function transforms precisely as that of a spin-2 primary operator. By definition, we have
\begin{equation} \langle T_{\mu\nu}(x_1)T_{\lambda\rho}(x_2) \rangle_{\mathcal{C}_A^{(n)}}^c=\langle T_{\mu\nu}(x_1)T_{\lambda\rho}(x_2) \rangle_{{\mathcal{C}_A^{(n)}}}-\langle T_{\mu\nu}(x_1) \rangle_{\mathcal{C}_A^{(n)}}\, \langle T_{\lambda\rho}(x_2) \rangle_{\mathcal{C}_A^{(n)}} \,.
\end{equation}
The second term on the right hand side of this equation contributes to $h_n^2$ order to the OPE coefficients and to $h_n^4$ order to the R\'{e}nyi mutual information. Hence this term does not contribute to the mutual information. In summary, we can safely conclude that the mutual information for various modes of the stress tensor can be obtained by simply setting $\Delta=d$ from the corresponding results for a primary spin-2 operator. As a result, we can claim that our bulk results from the graviton perfectly matches with the  contributions of the stress tensor to the MI in the boundary CFT.

The last remark is on the absence of the van Dam-Veltman-Zakharov discontinuity from the holographic mutual information.  In the above, we showed that the agreement between the bulk massless graviton and the boundary stress tensor. Actually, this agreement extends to the massive graviton and the corresponding spin-2 operator, as shown in Appendix A.2. From the field theory point of view, the dependence of the mutual information on the scaling dimension of the tensor operator is continuous. On the bulk side, we obtain the exact agreement with the boundary result is remarkable, suggesting that the massless limit of the graviton is well-defined and the van Dam-Veltman-Zakharov discontinuity is absent\cite{Kogan2000uy,Porrati2000cp}.

\subsection{Fermions}

Now, we study the  contribution from the fermionic field to the bulk mutual information .
In $AdS_{d+1}$, the Dirac matices $\Gamma$ have dimension
$N=trI_{d+1}=2^{[\frac{d+1}{2}]}$,
where ${[\frac{d+1}{2}]}$ equals to the largest integer that is smaller than ${\frac{d+1}{2}}$.
We choose the veilbein to be $e_{M }^a={\delta _{M }^a}/{z}$. The Dirac matrices in the tangent space satisfy $\{\gamma_a, \gamma_b\}=2\delta_{ab}$. The Gamma matrices in a curved space are defined by
\begin{equation}
\Gamma _{M }=\gamma _a e_{M }^a.
\end{equation}

Dual to a fermionic operator of dimension $\D$, there is a massive fermion in the bulk with mass $m=\D-d/2$.
The fermion propagator in the Euclidean AdS reads \cite{HS1998,fermion1999}
\begin{equation}
S(z,w)=-\sqrt{\frac{1}{w_0 z_0}} \left(G_{\Delta _-}(u) \left(\mathcal{P}_- \gamma _{\mu } z^{\mu }-\mathcal{P}_+ \gamma _{\mu } w^{\mu }\right)+G_{\Delta _+}(u) \left(\mathcal{P}_+ \gamma _{\mu } z^{\mu }-\mathcal{P}_- \gamma _{\mu } w^{\mu }\right)\right)\,,
\end{equation}
where $G_{\Delta\pm}(u)$ is the scalar propagator given in (\ref{scalarpropagator}) and
\begin{eqnarray}
\mathcal{P}_{\pm }=\frac{1}{2} \left(1\pm \gamma _d\right),
\Delta _{\pm }=\frac{d}{2}+\left(m\pm \frac{1}{2}\right).
\end{eqnarray}

in the large distance limit, we find
\begin{equation}
S(z,w)\overset{u\to \infty }{\approx }\mathcal{A} \left(I_N \left(w_d+z_d\right) -\gamma _{\mu } (z-w)^{\mu }+\gamma _i\gamma _0 \left(z^i-w^i\right)\right)\,,
\end{equation}
where
\be \mathcal{A}=-\frac{\Gamma \left(\frac{d+1}{2}+m\right)}{\left(\pi ^{d/2} 2^{\frac{d+3}{2}+m}\right) \Gamma \left(m+\frac{1}{2}\right) \sqrt{w_d z_d} u^{\frac{d+1}{2}+m}}\,.\ee
The gravity dual of the boundary spin-1 operator constructed from a fermionic operator is
 \begin{equation}\label{eq:o1}
O_M^{(f)(jj')}(r)=\bar{\psi }^{(j)}(r) \Gamma _M \psi ^{(j')}(r)-\psi ^{(j)}(r) \Gamma _M \bar{\psi }^{(j')}(r)\,.
\end{equation}
Here superscript $(f)$ stands for the operators constructed from the fermionic operators. All the other bilinear operators have vanishing contributions to the mutual information due to the antisymmetry of their indices\cite{CCHL2017}. However, just like previous cases, only the time-time component of the two-point function is relevant for our purpose. We find in the large distance limit
\begin{equation}
\left\langle O_0^{(f)(jj')}\left(r_A\right) O_0^{(f)(jj')}\left(r_B\right)\right\rangle\simeq
16\mathcal{A}^2N \frac{(2z_A z)^{2m+d-1}}{(x_A - x_B)^{4m+2d}}\,.
\end{equation}
Using the $1/n$ prescription, we compute the OPE coefficient as
\begin{equation}
C^{(A)0}_{(jj')}= (-1)^{j+j'} \frac{2^{-m-(d+3)/2}(R_{A})^{2m+d}
\sin^{-2m-d}\frac{\theta_{jj'}}{2n}}{\mathcal{A}\,N(z_{A})^{2m+d-1}}+O(n-1)\,.
\end{equation}
However, it is worth emphasizing that there are some subtleties when using the $1/n$ prescription for the fermions. First, the fermion propagator in the conifold geometry $C_A^{(n)}$ satisfies the boundary condition $G_F^{(n)}(\theta+2\pi n)=(-1)^{(n-1)}G_F^{(n)}(\theta)$, which is only periodic when the replica parameter $n$ is odd. We expect that odd $n$ result is already enough to derive the OPE coefficients. Second, we have included a factor $(-1)^{j+j'}$ in the OPE coefficient. This is correct since there will be a factor $(-1)$ for a fermion when it rotates $2\pi$\cite{Herzog2015}.

At last we find the fermion contribution to the bulk mutual information
\begin{equation}
I(A,B)|_{s=1}=\frac{\sqrt{\pi } 2^{\left\lfloor \frac{d+1}{2}\right\rfloor -2} \Gamma (d+2 m+1)}{4^{d+2 m} \Gamma \left(d+2 m+\frac{3}{2}\right)}z_{cr}^{d+2m}.
\end{equation}
Note that it is positive which is different in sign from the vector operators constructed from the scalar operator. This is reasonable since it gives the leading order contribution to the mutual information.

For $d$ is odd, the result is the same as that of a Dirac fermion in the boundary CFT\cite{CCHL2017}. However, for $d$ is even, it is just one half of that. This can be easily understood in the AdS/CFT correspondence. The bulk fermion has different duals in the boundary in different dimensions. When $d$ is odd, it is dual to a Dirac fermion operator in the boundary.
However, when $d$ is even, it corresponds to a Weyl fermion operator, which can be viewed as one half of a Dirac fermion \cite{fermion1998,fermion1999_2,Iqbal2009}. We can write a Dirac fermion $\Psi=\Psi_1 +\Psi_2$ in which $\Psi_{1,2}$ are Weyl fermions. For a general primary field $O$, its contribution to the mutual information $I\sim\frac{\left\langle O\right\rangle \left\langle O\right\rangle }{\left\langle OO\right\rangle }$. Two independent fields $O_1$ and $O_2$ contribute as $I\sim\frac{\left\langle O_1\right\rangle \left\langle O_1\right\rangle }{\left\langle O_1 O_1\right\rangle }+\frac{\left\langle O_2\right\rangle \left\langle O_2\right\rangle }{\left\langle O_2 O_2\right\rangle }$. Since the bulk fermion is dual to only a Weyl fermion (half of a Dirac fermion) on the boundary when $d$ is even, its contribution to the bulk MI is only half of that contributed by a boundary Dirac fermion. In short, from the AdS/CFT correspondence, the bulk MI from the Dirac fermions in different dimensions are in match with the boundary results.

\section{Conclusions}

In this paper, we tried to understand holographically the universal behaviors in the leading orders of the mutual information between two disjoint spheres in a CFT. Such universal behaviors have been found in \cite{CCHL2017} by using the operator product expansion of the twist operator and the $1/n$ prescription. Holographically, the spherical
twist operator can be understood as the non-local hemisphere.
In the large distance regime, we can still use the operator product expansion of the hemisphere to simplify the calculations. As we are interested in the mutual information, we can safely ignore the backreaction of the twist operator to the geometry. Effectively we can still applied the replica trick in the bulk without worrying about the backreaction. Moreover in the $n\to 1$ limit, the fields could be treated as the generalized free theory such that the interaction can be ignored.  In the bulk computation, the fields are treated as the free field as well. Therefore in the computation of the holographic mutual information, we consider the free fields in a fixed background. Especially to compute the OPE coefficients, we could focus on the free fields in a space with conical singularity such that we may apply the $1/n$ prescription to read the coefficients. By explicit computation, we showed that the leading mutual information in a CFT, no matter what kind of operator leads to, the scalar, the vector, the tensor or the fermionic type, can be reproduced from the holographic computation of the dual corresponding field. %In this sense, this is similar to the computation of 1-loop determinant of the fields.

In retrospect, the universal behaviors in the leading mutual information in a general CFT suggests that it is independent of the details of the AdS/CFT correspondence, namely the explicit construction of the AdS gravity and the dual CFT. Such behaviors relies only on the symmetry. From the general lesson in the AdS/CFT correspondence, the fields in AdS  is dual to the operator in the boundary theory. In other words, the behavior of the free fields in AdS could be captured by the dual operator constrained by the conformal symmetry, and vice versa. Our study
gives another piece of evidence to support this picture, though in a subtler way.
%The mutual information can then be calculated with the help of $1/n$ prescription. Since this framework is applicable to quantum fields in curved space, we can use it to calculate the quantum corrections to the holographic mutual information.

%We then reviewed the calculations of the leading order contributions to the holographic MI of the bulk scalar field and generalized the results to sub-leading orders. To compare with the results in boundary CFT, we constructed the bilinear bulk fields dual to the boundary bilinear operators. We find we have to apply a projection for dual fields in the bulk in order to get the same results as in CFT.

Even though the conclusion might not be a big surprise, the procedure to get this picture is remarkable. The leading contribution is from the bilinear operator composed of the fields at different replicas. In the scalar field case, we could even discuss the next-to-leading order contribution, which is from the bilinear operator with a derivative. In this case, we found a new form of the projection operator defined on the slice of fixed radius. It was defined to peel off the radial components so as to make the operators in the same form as in the boundary CFT. In the gauge field case, we can treat this as a particular gauge choice. However, when considering the massive fields, there is no such understanding. We wish the construction could be useful in other situations.

Another remarkable point is on the gauge fields. For the fields with gauge symmetry,
 including the massless vector bosons and the massless gravitons, we found that the gauge parts in the propagators played an indispensable role in the calculation, even though the final results are gauge independent. We argue that this is due to the gauge symmetry breaking around the entangling surfaces. It gives rise to extra physical degrees of freedom to contribute to the MI. In fact, we also calculated the MI from the massive bulk fields in Appendix A. The results match exactly with boundary results of the vector and tensor operator with generic scaling dimensions. This supports our arguments since a massive vector field has one more degrees of freedom than  the gauge boson. As a byproduct, we showed that the absence of the van Dam-Veltman-Zakharov discontinuity in the computation of the holographic mutual information.

%We also calculated the contribution of  Dirac fermion fields to the holographic MI.
%For $d=odd$, the result is the same as that of Dirac fermion in CFT. For $d=even$, it is just one half of that. This   matches exactly with what AdS/CFT correspondence says. For $d=odd$, a bulk Dirac fermion is dual to a boundary Dirac fermion operator. However it is dual to a Weyl fermion operator when $d=even$, which can be view as one half of Dirac fermion.

In our calculations, we treated AdS spacetime as a background. As a result, our results are meaningful only when we take $n\to 1$, so that the twist operator  can be treated as a probe. Thus, it seems impossible to generalize the discussion to the R\'enyi entropy, since in this case the twist operator is heavy and would affect the background spacetime significantly. The key problem we really face with is how to expand the twist (or surface) operator. There are other methods to construct the gravity dual of a surface operator in CFT, like ``bubbling" surface operator as mentioned in \cite{DGM}, which have taken into account of the back reaction so that it can be used to get quantum correction to the R\'enyi entropy. It would be interesting to investigate this possibility. %It may be a clue to perfect the dictionary of AdS/CFT correspondence.

In \cite{CCHL2017}, it was shown that the mutual information could be expanded in terms of the conformal blocks. The conformal block carries the higher order contribution of the cross ratio. In the free fermion case, the conformal block expansion fits better with the numerical study than the simple leading order expansion of the conformal block. It would be interesting to see if one can find the conformal block expansion in the holographic picture.

\section*{Acknowledgments}

We would like to thank Jiang Long, Lin Chen and Peng-Xiang Hao for stimulating discussions. B. Chen would like to thank the participants of the advanced workshop ``Dark Energy and Fundamental Theory" supported by the Special Fund for Theoretical Physics from the National Natural Science Foundations of China with Grant No. 11447613 for stimulating discussion. C.-Y. Zhang thanks the APCTP focus program "Geometry and holography for quantum criticality" in Pohang, Korea. Z.-Y. Fan is supported in part by NSFC Grants No. 11273009 and No. 11303006 and also supported by Guangdong Innovation Team for Astrophysics(2014KCXTD014). C.-Y. Zhang is supported by National Postdoctoral Program for Innovative Talents BX201600005. B. Chen and W.-M. Li were supported in part by NSFC Grant No.~11275010, No.~11325522, No.~11335012 and No. 11735001.

\appendix
\section{ Massive Field}
We consider the holographic mutual information from the massive vector and tensor fields in this appendix. We show that their contributions  are well matched with the boundary results of the spin-1 and spin-2 primary operators.

\subsection{Massive Vector Fields}

We compute the holographic MI contributed by the massive vector field in this part.
For a massive vector field, the propagator is\cite{Massive2014}
\begin{align}
\left\langle A_{M}(x)A_{N'}(x')\right\rangle & =G_{MN'}(x,x')=-(\partial_{M}\partial_{N'}u)g_{0}(u)+\left(\partial_{M}u\partial_{N'}u\right)g_{1}(u),\\
g_{0}(u) & =(d-\Delta)F_{1}(u)-\frac{1+u}{u}F_{2}(u)\nonumber, \\
g_{1}(u) & =\frac{(1+u)(d-\Delta)}{u(2+u)}F_{1}(u)-\frac{d+(1+u)^{2}}{u^{2}(2+u)}F_{2}(u)\nonumber, \\
F_{1}(u) & =\mathcal{N}(2u)^{-\Delta}{}_{2}F_{1}\left(\Delta,\frac{1-d+2\Delta}{2},1-d+2\Delta,-\frac{2}{u}\right)\nonumber ,\\
F_{2}(u) & =\mathcal{N}(2u)^{-\Delta}{}_{2}F_{1}\left(\Delta+1,\frac{1-d+2\Delta}{2},1-d+2\Delta,-\frac{2}{u}\right)\nonumber ,\\
\mathcal{N} & =\frac{\Gamma(\Delta+1)}{2\pi^{d/2}(d-1-\Delta)(\Delta+1)\Gamma(\Delta+1-d/2)}\nonumber.
\end{align}
The specifical form of the normalization constant $\mathcal{N}$ is irrelevant because it is cancelled in the final results.

The two operators having the leading order contributions to the holographic MI can be defined in the same forms as (\ref{O0O1}).
For the spin-0 case, its contribution to the MI can be formally written as
\begin{equation}
I(A,B)|_{s=0}=\frac{n}{2 (n-1)}\sum _{j=1}^{n-1} C^A_{(0 j)} C^B_{(0 j)} \left\langle O^{(v)(0 j)}\left(r_A\right) O^{(v)(0 j)}\left(r_B\right)\right\rangle.
\end{equation}
The two-point function in the large separation is
\begin{align}
\left\langle O^{(v)(jj')}(r_{A})O^{(v)(jj')}(r_{B})\right\rangle _{M}\simeq  \mathcal{N}^2d(d-1-\Delta)^{2}\frac{(2z_{A}z_{B})^{2\Delta}}{|x_{A}-x_{B}|{}^{4\Delta}}.
\end{align}
Note that it can not be reduced to the massless gauge fields  by simply taking limit $\Delta \to d-1$.
Using the $1/n$ prescription, we can get
\begin{equation}
C_{(0j)}^{A}=\frac{2^{-\Delta}}{d(d-1-\Delta)}z_{A}^{-2\Delta}R_{A}^{2\Delta}\frac{[(d-1)n^{2}-1]\sin^{-2\Delta}\left(\frac{\theta_{j}}{2n}\right)}{n^{2}}.
\end{equation}
It follows that the corresponding mutual information is
\begin{equation}\label{eq:mv0}
I(A,B)|_{s=0}=\frac{(d-2)^{2}}{d}\frac{\sqrt{\pi}}{4^{2\Delta+1}}\frac{\Gamma(2\Delta+1)}{\Gamma(2\Delta+\frac{3}{2})}z_{cr}^{2\Delta}
,\end{equation}
which matches with (\ref{vector0}).

For the spin-2 operator, the contribution to MI is
\begin{equation}\label{eq:mv2}
I(A,B)|_{s=2}=\frac{n}{2 (n-1)}\sum _{j=0}^{n-1} C^{(A)00}_{(0 j)} C^{(B)00}_{(0 j)} \left\langle O_{00}^{(v)(0j)}\left(r_A\right) O_{00}^{(v)(0j)}\left(r_B\right)\right\rangle.
\end{equation}
The time-time component of the two-point function at the large separation is
\begin{equation}
\left\langle O_{00}^{(v)(jj')}\left(r_A\right) O_{00}^{(v)(jj')}\left(r_B\right)\right\rangle \simeq \frac{(d-1) \mathcal{N}^2 (d-\Delta -1)^2}{d}\frac{(z_{A}z_{B})^{2\Delta-2}}{|x_{A}-x_{B}|{}^{4\Delta}}.
\end{equation}
The OPE coefficient $C^{(A)00}_{(jj')}$ can be calculated as
\begin{equation}
C_{(0j)}^{(A)00}=\frac{-2^{-\Delta+2}}{(d-1-\Delta)}\frac{(1+n^{2})}{n^{2}}(z_{A})^{-(2\Delta-2)}R_{A}^{2\Delta}\sin^{-2\Delta}\left(\frac{\theta_{j}}{2n}\right).
\end{equation}
Finally, we get
\begin{align}
I(A,B)|_{s=2}= & \frac{d-1}{d}\frac{\sqrt{\pi}}{4^{2\Delta}}\frac{\Gamma(2\Delta+1)}{\Gamma(2\Delta+\frac{3}{2})}z_{cr}^{2\Delta}
,\end{align}
which agrees with (\ref{vector2}).
%These results are exactly the same as those of a spin-1 primary operator living in Euclidean CFT$_d$ \cite{CCHL2017}.

\subsection{Massive Tensor Field}

Now we calculate the contributions of the symmetric and traceless massive spin-2 tensor field $\hat{G}_{MN}$ to the holographic MI.  The propagator is given by\cite{Massive1999,Massive2014}
\begin{align}
\left\langle \hat{G}_{MN}(x_{A})\hat{G}_{EF}(x_{B})\right\rangle _{M}= & G_{MN,EF}(x_{A},x_{B})=\sum_{i=1}^{5}A^{(i)}T_{MN,EF}^{(i)},\\
T_{MN,EF}^{(1)}= & h_{MN}h_{EF}\nonumber, \\
T_{MN,EF}^{(2)}= & (\partial_{M}u\partial_{N}u)(\partial_{E}u\partial_{F}u)\nonumber, \\
T_{MN,EF}^{(3)}= & (\partial_{M}\partial_{E}u)(\partial_{N}\partial_{F}u)+(\partial_{M}\partial_{F}u)(\partial_{N}\partial_{E}u)\nonumber ,\\
T_{MN,EF}^{(4)}= & h_{MN}(\partial_{E}u\partial_{F}u)+h_{EF}(\partial_{M}u\partial_{N}u)\nonumber ,\\
T_{MN,EF}^{(5)}= & (\partial_{M}\partial_{E}u)(\partial_{N}u\partial_{F}u)+(\partial_{N}\partial_{E}u)(\partial_{M}u\partial_{F}u)\nonumber \\
 & +(\partial_{M}\partial_{F}u)(\partial_{N}u\partial_{E}u)+(\partial_{N}\partial_{F}u)(\partial_{M}u\partial_{E}u)\nonumber
.\end{align}
The coefficients are specified by
\begin{align}
A^{(1)}= & -\frac{1+d-u(2+u)}{(1+d)^{2}}g_{0}-\frac{u(1+u)(u+2)}{(1+d)^{2}}g_{1}+\frac{u^{2}(2+u)^{2}}{(1+d)^{2}}g_{2}\nonumber ,\\
A^{(2)}= & g_{2},
A^{(3)}=  \frac{1}{2}g_{0}\nonumber, A^{(5)}=  -\frac{1}{4}g_{1} \\
A^{(4)}= & -\frac{1}{1+d}g_{0}+\frac{1+u}{1+d}g_{1}-\frac{u(u+2)}{1+d}g_{2}\nonumber ,
\nonumber,
\end{align}
where in the large distance limit,
\begin{align}
g_{k}\approx& \frac{J!}{k!(J-k)!}u^{-\Delta-k}.
\end{align}
The relevant operators having contributions to the MI have the same form defined in (\ref{O21}-\ref{O4}).

The calculations are straightforward, as illustrated for the gravitons. For the spin-0 case, we get
\begin{align}\label{eq:mt0}
I(A, B)|_{s=0}&=\frac{n}{2 (n-1)}\sum _{j=1}^{n-1} C^A_{(0 j)} C^B_{(0j)} \left\langle O^{(t)(0j)}\left(r_A\right) O^{(t)(0j)}\left(r_B\right)\right\rangle\nonumber\\
&=\frac{(d-2)^{2}(d-1)}{2(d+2)}\frac{1}{4^{2\Delta+1}}
\frac{\sqrt{\pi}\Gamma(2\Delta+1)}{\Gamma(2\Delta+\frac{3}{2})}z_{cr}^{2\Delta}
.\end{align}
For the spin-2 case, we get
\begin{align}\label{eq:mt2}
I(A, B)|_{s=2}&=\frac{n}{2 (n-1)}\sum _{j=0}^{n-1}C^{(A)00}_{(0 j)} C^{(B)00}_{(0 j)} \left\langle O_{00}^{(t)(0j)}\left(r_A\right) O_{00}^{(t)(0j)}\left(r_B\right)\right\rangle\nonumber\\
&=\frac{(d-1)(d-2)}{(d+4)}\frac{\sqrt{\pi}\Gamma(2\Delta+1)}
{4^{2\Delta}\Gamma(2\Delta+\frac{3}{2})}z_{cr}^{2\Delta}
.\end{align}
For the spin-4 operator, the contribution to the MI is
\begin{equation}
I(A, B)|_{s=4}=\frac{n}{2 (n-1)}\sum _{j=1}^{n-1}  C^{(A)0000}_{(0j)} C^{(B)0000}_{(0j)}  \left\langle O_{0000}^{(t)(0j)}\left(r_A\right) O_{0000}^{(t)(0j)}\left(r_B\right)\right\rangle,
\end{equation}
where the OPE coefficient $C^{(A)0000}_{(jj')}$ can be calculated as
\begin{equation}
C^{(A)0000}_{(jj')}=\frac{\left\langle O_{0000}^{(t)(jj')}(r)\right\rangle {}_{C_A^n}}{\left\langle O_{0000}^{(t)(jj')}(r) O_{0000}^{(t)(jj')}\left(r_A\right)\right\rangle }.
\end{equation}
Given the two-point function of the operator in large separation
\begin{equation}
\left\langle O_{0000}^{(t)(jj')}\left(r_A\right) O_{0000}^{(t)(jj')}\left(r_B\right)\right\rangle \simeq\frac{(d-1)  (d+1))}{(d+4) (d+2)}\frac{(\Delta +1)^2 s_0^2}{(\Delta -1)^2}\frac{\left(z_A z_B\right){}^{2 \Delta -4}}{\left|x_A-x_B|^{4 \Delta }\right.}.
\end{equation}
and using $1/n$ prescription, we can easily obtain $C^{(A)0000}_{(jj')}$. Finally, we deduce
\begin{equation}\label{eq:mt4}
I(A, B)|_{s=4}=\frac{(d-1) (d+1)}{(d+4) (d+2) 4^{2 \Delta -1}}\frac{\sqrt{\pi } \Gamma (2 \Delta +1)}{\Gamma \left(2 \Delta +\frac{3}{2}\right)}z_{cr}^{2\Delta}.
\end{equation}
All these results are exactly the same as those from a primary spin-2 operators in the Euclidean CFT$_d$, see (\ref{tensor4}, \ref{tensor2}, \ref{tensor0}).

\section{Graviton propagator}

The graviton propagator in $AdS_{D}$ in the Landau gauge\cite{Graviton1999_1} can be written as (\ref{eq:gra}). In this formula, the bilocal tensor structure $O_{MN,EF}$ is given by
\begin{align}
O_{MN,E'F'}^{(1)}= & g_{MN}g_{E'F'},\nonumber \\
O_{MN,E'F'}^{(2)}= & n_{M}n_{N}n_{E'}n_{F'},\nonumber \\
O_{MN,E'F'}^{(3)}= & g_{ME'}g_{NF'}+g_{MF'}g_{NE'},\nonumber \\
O_{MN,E'F'}^{(4)}= & g_{MN}n_{E'}n_{F'}+g_{E'F'}n_{M}n_{N},\nonumber \\
O_{MN,E'F'}^{(5)}= & g_{ME'}n_{N}n_{F'}+g_{MF'}n_{N}n_{E'}+g_{NE'}n_{M}n_{F'}+g_{NF'}n_{M}n_{E'}\,,
\end{align}
in which
\begin{eqnarray}
n_{M}&= & \frac{1}{\sqrt{u(u+2)}}\partial_{M}u,\nonumber \\
g_{ME'}&= & \partial_{M}\partial_{E'}u+\frac{\partial_{M}u\partial_{E'}u}{u+2}.
\end{eqnarray}
The coefficients $T$ and $G$ are respectively,
\begin{eqnarray}
T&= & -\frac{1}{(D-2)(D-1)}\frac{\Gamma(D)\Gamma(\frac{D}{2}+1)}{\Gamma(D+2)\pi^{\frac{D}{2}}2^{D}}\frac{1}{x^{D}}F(D,\frac{D}{2}+1,D+2,\frac{1}{x})\,,\nonumber\\
G^{(1)}&= & \frac{1}{D(D-2)}\left[\frac{4(x-1)^{2}x^{2}}{D+1}g''+4x(x-1)(2x-1)g'+\left(4Dx(x-1)+D-2\right)g\right]\,,\nonumber \\
G^{(2)}&= & -4(x-1)^{2}\left[\frac{x^{2}}{D(D+1)}g''+\frac{2(D+2)x}{D(D+1)}g'+g\right]\nonumber\,, \\
G^{(3)}&= & -\frac{1}{2(D-2)}\left[\frac{4(D-1)x^{2}(x-1)^{2}}{D(D+1)}g''+\frac{4(D-1)x(x-1)(2x-1)}{D}g'\right]\nonumber\\
&&-\frac{1}{2(D-2)}\left[\left(4(D-1)x(x-1)+D-2\right)g\right]\,,\nonumber \\
G^{(4)}&= & -\frac{4x(x-1)}{D(D-2)}\left[\frac{x(x-1)}{D+1}g''+(2x-1)g'+Dg\right]\nonumber \\
G^{(5)}&= & -\frac{x-1}{D-2}\left[\frac{2(D-1)x^{2}(x-1)}{D(D+1)}g''+\frac{x(4(D-1)x-3D+4)}{D}g'\right]\\
&&-\frac{x-1}{D-2}\left[\left(2(D-1)x-D+2\right)g\right]\,, \nonumber
\end{eqnarray}
where $x=\frac{u+2}{2}$ and
\begin{eqnarray}
g&=& g_{part}^{0}(x)+g_{part}^{1}(x)+\frac{D(D+1)\Gamma(\frac{D}{2}+1)}{(D-2)(D-1)2^{D+2}\pi^{D/2}}
g_{1}\nonumber\\
&&+(-1)^{\frac{D-1}{2}}\frac{\Gamma(\frac{D+3}{2})D}
{(D+2)(D-1)^{2}\pi^{\frac{D-1}{2}}}g_{2}.
\end{eqnarray}
In the above formula,
 \begin{eqnarray}
 g_{part}^{0}(x)&= & \frac{D(D+1)\Gamma(\frac{D}{2}-1)}{(D-1)2^{D+2}\pi^{D/2}}
 \frac{d}{dx}\frac{\int_{0}^{x}dx'F(-D+2,1,-\frac{D}{2}+1,x')}{[x(x-1)]^{D/2}},\nonumber \\
g_{part}^{1}(x)&= & -\frac{D}{(D-2)(D-1)}\frac{(D+1)\Gamma(\frac{D}{2}+1)}{2^{D+2}D\pi^{D/2}}
\frac{4(D-1)x(x-1)+D}{[x(x-1)]^{D/2+1}},\nonumber \\
g_{1}&= & \frac{2x-1}{[x(x-1)]^{D/2+1}},\nonumber \\
g_{2}&= & F(2,D+1,\frac{D}{2}+2,1-x).\nonumber
 \end{eqnarray}
 In the large distance limit,
\begin{equation}\label{eq:bn}
g\rightarrow  f(D)x^{-D}-b(D)x^{-D-1}\ln x\,,
\end{equation}
  where 
\begin{align}
f(D)=-\frac{(D+1)}{(D-2)}\frac{\Gamma(\frac{D}{2}+1)}{2^{D}\pi^{D/2}}, & \hspace{3ex}b(D)=(-1)^{\frac{D+1}{2}}\frac{(D+1)D}{(D-1)^{2}}\frac{\Gamma(\frac{D}{2}+1)}{2^{D+1}\pi^{\frac{D}{2}}}.
\end{align}

\end{document}